\newcommand{\appropto}{\mathrel{\vcenter{
  \offinterlineskip\halign{\hfil$##$\cr
    \propto\cr\noalign{\kern2pt}\sim\cr\noalign{\kern-2pt}}}}}

\documentclass[twocolumn]{aastex62}

\received{6 Oct 2017}
\revised{12 Oct 2018}
\accepted{14 Oct 2018}

\shorttitle{A Role for Turbulence in Circumgalactic Precipitation}
\shortauthors{Voit}

\begin{document}

\title{\bf A Role for Turbulence in Circumgalactic Precipitation}

\author{G.\ Mark Voit}  
\affiliation{Department of Physics and Astronomy,
                 Michigan State University,
                 East Lansing, MI 48824}

\begin{abstract}
The cooling time $t_{\rm cool}$ of the hot ambient medium pervading a massive galaxy does not drop much below 10 times the freefall time $t_{\rm ff}$ at any radius.  Theoretical models have accounted for this finding by hypothesizing that cold clouds start to condense out of the ambient medium when $t_{\rm cool} / t_{\rm ff} \lesssim 10$ and fuel a strong black hole feedback response, but they have not yet provided a simple explanation for the critical $t_{\rm cool} / t_{\rm ff}$ ratio.  This paper explores a heuristic model for condensation linking the critical ratio to turbulent driving of gravity-wave oscillations.  In the linear regime, internal gravity waves are thermally unstable in a thermally balanced medium.  Buoyancy oscillations in a balanced medium with $t_{\rm cool} / t_{\rm ff} \gg 1$ therefore grow until they saturate without condensing at an amplitude depending on $t_{\rm cool} / t_{\rm ff}$.  However, in a medium with $10 \lesssim t_{\rm cool} / t_{\rm ff} \lesssim 20$, turbulence with a velocity dispersion roughly half the galaxy's stellar velocity dispersion can drive those oscillations into condensation.  Intriguingly, this is indeed the gas-phase velocity dispersion observed among multiphase galaxy cluster cores.  It is therefore possible that both the critical $t_{\rm cool} / t_{\rm ff}$ ratio for condensation of ambient gas and the level of turbulence in that gas are determined by coupling between condensation, feedback, and turbulence.  Such a system can converge to a well-regulated equilibrium state, if the fraction of feedback energy going into turbulence is subdominant.
\end{abstract}

\keywords{galaxies: clusters: intracluster medium --  galaxies: groups: general -- galaxies: ISM}

\section{Introduction}
\setcounter{footnote}{0}

Not long ago, \citet{Voit_2017_BigPaper} presented a global model for the hot atmospheres of galaxies.  In that model the thermodynamic state of the ambient gas surrounding a galaxy is regulated by a phenomenon we call precipitation.   The proposed regulation mechanism depends on a critical assumption, that the transition from a single-phase hot ambient medium to a multiphase medium dramatically boosts the energy input from feedback into the circumgalactic medium, as cold clouds condense out of the hot gas.  Those clouds close the feedback loop by raining down into the galaxy's center and fueling the central engine through a ``cold feedback" process called ``chaotic cold accretion" \citep[see, e.g.,][]{ps05,ps10,McCourt+2012MNRAS.419.3319M,Sharma_2012MNRAS.420.3174S,Gaspari+2012ApJ...746...94G,Gaspari+2013MNRAS.432.3401G,Gaspari_2017MNRAS.466..677G}.  Such a feedback loop tends to suspend the circumgalactic medium at the phase-transition threshold as long as certain conditions are satisfied.  The most critical condition is that feedback heating of the circumgalactic medium must not be too centrally concentrated, because feedback then stimulates convection and runaway thermal instability that fails to self-regulate.  Instead, heat must be introduced into the circumgalactic medium in a manner that preserves a large-scale radial gradient in specific entropy, because the entropy gradient maintains convective stability and limits the condensation rate \citep[e.g.,][]{Meece_2017ApJ...841..133M}.

These ideas have a long history that was extensively discussed by both \citet{Voit_2017_BigPaper} and \citet{Gaspari_2017MNRAS.466..677G}.  Readers interested in more complete discussions can consult those papers and the citations within them.  The scope of this particular paper is narrower.  It focuses exclusively on how turbulence alters the global model presented by \citet{Voit_2017_BigPaper}.  That paper advocated a structure for the circumgalactic medium with an inner ``isentropic" zone and an outer ``power-law" zone. In the power-law zone the entropy index $K = kT n_e^{-2/3}$, defined with respect to electron density $n_e$ in a medium with an adiabatic index of 5/3, rises as a power-law function of radius $r$.  If the cooling time $t_{\rm cool}$ required for the gas to radiate $3kT/2$ per particle significantly exceeds the freefall time $t_{\rm ff} \equiv \sqrt{2 r^2 / v_c^2}$ in a gravitational potential with circular velocity $v_c$ at radius $r$, then buoyancy tends to suppress condensation in the power-law zone.  Without turbulence, condensation proceeds through two channels: (1) thermal instability in the isentropic zone, in which buoyancy cannot damp condensation, and (2) uplift of low-entropy gas into the power-law zone by outflows emanating from the central region \citep[see also][]{LiBryan2014ApJ...789..153L,McNamara_2016ApJ...830...79M}.

However, observations suggest the presence of a third zone in between the other two.  \citet{Hogan_2017_tctff} in particular have drawn attention to the fact that the central entropy profiles in multiphase galaxy cluster cores are not flat, except perhaps at radii much smaller than the size of the multiphase region, which can span $\gtrsim 20$~kpc.  At radii from $\sim 5$ to 20 kpc in the multiphase cores of many galaxy clusters, the ambient $t_{\rm cool} / t_{\rm ff}$ ratio appears to be a nearly constant function of radius, with values in the range $10 \lesssim t_{\rm cool} / t_{\rm ff} \lesssim 20$.  Similar results were found for individual massive elliptical galaxies by \citet{Voit+2015ApJ...803L..21V}, and both of those findings are in alignment with the floor seen at $t_{\rm cool} / t_{\rm ff} \approx 10$ by \citet{Voit_2015Natur.519..203V} in the galaxy cluster population as a whole \citep[see also][]{McCourt+2012MNRAS.419.3319M,Sharma+2012MNRAS.427.1219S,VoitDonahue2015ApJ...799L...1V}.  Furthermore, the $K \propto r^{2/3}$ entropy profile corresponding to constant $t_{\rm cool} / t_{\rm ff}$ in an isothermal potential (i.e. $t_{\rm ff} \propto r$) agrees with the median inner entropy slope found in large samples of central galaxies by \citet{Panagoulia_2014MNRAS.438.2341P}, \citet{Hogan_2017_tctff}, and \citet{Sanders_2017arXiv170509299S}.

This paper shows that turbulence is capable of stimulating condensation in circumgalactic gas with $K \propto r^{2/3}$ and $10 \lesssim t_{\rm cool} / t_{\rm ff} \lesssim 20$, as long as it is strong enough to counteract the tendency for buoyancy to suppress condensation.  The required turbulent velocity dispersion is $\sigma_{\rm t} \approx 0.5 \, \sigma_v$, where $\sigma_v$ is the stellar velocity dispersion of the central galaxy.  Notably, this is similar to the velocity dispersion observed among molecular clouds in multiphase galaxy cluster cores \citep{McNamara_2014ApJ...785...44M,Russell_2016MNRAS.458.3134R}, in the X-ray--emitting gas at the center of the Perseus Cluster \citep{Hitomi_Perseus_2016Natur.535..117H}, and also among Mg II absorption-line systems in the circumgalactic media of a much larger sample of luminous early-type galaxies \citep{Huang_2016MNRAS.455.1713H}.  Section 2 of the paper recaps the key results from \citet{Voit_2017_BigPaper}, which reviewed the relevant literature on thermal instability and added some new insights.  Section 3 builds on those results to construct a heuristic model that links condensation with turbulence.  Section 4 outlines how coupling between precipitation, feedback, and turbulence can form a self-regulating feedback loop that converges to an equilibrium state with $10 \lesssim t_{\rm cool} / t_{\rm ff} \lesssim 20$ and $\sigma_{\rm t} / \sigma_v \approx 0.5$.  Section 5 summarizes and reflects on the main findings.

\section{Thermally Unstable Internal Gravity Waves}
\label{sec-UnstableWaves}

The term ``thermal instability" in this paper refers to a Lagrangian entropy contrast with an amplitude that grows with time.  A thermally unstable gas parcel begins with entropy $K_1$, and the magnitude of its Lagrangian contrast, $\Delta_K \equiv \ln (K/K_1)$, either grows monotonically with time $t$ or oscillates with increasing amplitude.   In a thermally balanced medium with $t_{\rm cool} / t_{\rm ff} \gg 1$ and a significant entropy gradient, thermal instability is oscillatory because it drives internal gravity waves \citep[e.g.,][]{Defouw_1970ApJ...160..659D}.  For example, consider a spherically symmetric medium in which the mean entropy is $\bar{K} = K_1 (r/r_1)^{\alpha_K}$.  The Eulerian entropy contrast of an unstable disturbance is then $\delta_K \equiv \ln (K/\bar{K}) = \Delta_K - \alpha_K \Delta_r$, where $\Delta_r \equiv \ln (r/r_1)$.  If fluctuating pressure forces are unimportant, buoyancy causes an adiabatic disturbance ($\Delta_K = 0$) to oscillate as an internal gravity wave at the Brunt--V\"ais\"al\"a frequency $\omega_{\rm buoy} = (6\alpha_K/5)^{1/2} t_{\rm ff}^{-1}$.  Thermal instability tends to amplify those gravity waves if perturbations cool faster than their surroundings in places where $\delta_K < 0$ and more slowly in places where $\delta_K > 0$.  Amplification occurs because of a phase difference of magnitude $\sim t_{\rm ff} / t_{\rm cool}$ between $\delta_K$ and $- \alpha_K \Delta_r$ that allows some heating to happen during a period with $\Delta_r > 0$ and some cooling to happen during a period with $\Delta_r < 0$.  

The literature on the nonlinear fate of these seemingly simple oscillations and their implications for condensation is surprisingly complex and often confusing.\footnote{A more positive way to express the same thought would be to say that the physics of circumgalactic condensation turns out to be rich and interesting.}  \citet{Voit_2017_BigPaper} emphasized the following points while attempting to review and summarize the subject:
\begin{enumerate}

\item  A single-phase medium cannot develop persistent multiphase structure without some help from feedback.  The reason comes down to fundamental thermodynamics.  Development of multi-temperature gas in an idealized system with constant total mass and energy corresponds to a reduction in total entropy and requires a flow of free energy through the system.  Essentially, free energy released by feedback is needed to replace radiative energy losses from the ambient medium so that it remains hot and diffuse while lower-entropy clouds condense within it.  

\item  Without a heat source, thermal instability and cooling of the ambient medium occur on similar timescales, and that prevents persistent entropy contrasts from developing.  Most of the recent analyses of circumgalactic condensation have therefore assumed thermal balance on equipotential surfaces \citep[e.g.,][]{McCourt+2012MNRAS.419.3319M,Gaspari+2013MNRAS.432.3401G,Meece_2015ApJ...808...43M,ChoudhurySharma_2016MNRAS.457.2554C}, but the results are not much different for fluctuating heating rates, as long as the fluctuation timescale does not exceed the cooling time \citep[e.g.,][]{Sharma_2012MNRAS.420.3174S}.

\item  In a hydrostatic, thermally balanced medium that is also convectively stable, internal gravity waves are thermally unstable, while sound waves are dissipative.  The dispersion relation for thermally unstable gravity waves with amplitude $\propto e^{-i \omega t}$ is
\begin{equation}
  \omega^2 - i \omega_{\rm ti} \omega - \frac {k_\perp^2} {k^2} \omega_{\rm buoy}^2 = 0
  \; \; ,
 \label{eq-gmodes}
\end{equation}
where $\omega_{\rm ti} \approx t_{\rm cool}^{-1}$ is the rate at which thermal instability develops, $k$ is the wavenumber of the disturbance, $k_\perp$ is the wavenumber in the direction perpendicular to gravity, and short wavelengths ($kr \gg 1$) have been assumed \citep[e.g.,][]{Malagoli_1987ApJ...319..632M,bnf09}.  The dispersion relation for short-wavelength sound waves is
\begin{equation}
  \omega^2 + i \omega_{\rm sw} \omega - c_s^2 k^2  = 0
  \; \; ,
 \label{eq-pmodes}
\end{equation}
where $\omega_{\rm sw} \approx t_{\rm cool}^{-1}$ is the rate at which sound waves damp and $c_s$ is the sound speed.\footnote{Note that this definition of $\omega_{\rm sw}$ differs in sign from the definition of the same quantity in \citet{Voit_2017_BigPaper}.}  \\

\item  Numerical simulations \citep[e.g.,][]{McCourt+2012MNRAS.419.3319M,Meece_2015ApJ...808...43M,ChoudhurySharma_2016MNRAS.457.2554C} show that thermally unstable gravity waves in a hydrostatic and thermally balanced ambient medium saturate when the radial displacement amplitude approaches $\Delta_r \sim \omega_{\rm ti} / \omega_{\rm buoy}$.  The Eulerian entropy perturbation amplitude corresponding to this displacement is 
\begin{equation}
  | \delta_K | \approx \alpha_K ( \omega_{\rm ti} / \omega_{\rm buoy} ) 
  		     \approx \alpha_K^{1/2} (t_{\rm ff} / t_{\rm cool})
\end{equation}
as long as $\alpha_K \gg (t_{\rm ff} / t_{\rm cool})^2$. Reducing $\alpha_K$ consequently reduces the saturation amplitude for entropy perturbations until it reaches a minimum at $| \delta_K | \sim (t_{\rm ff} / t_{\rm cool})^2$ for $\alpha_K \sim (t_{\rm ff} / t_{\rm cool})^2$.  

\item  Saturation of oscillatory thermal instability occurs when the gravity wave amplitude grows large enough for nonlinear mode coupling to transfer gravity-wave energy into dissipative modes at a rate $\sim \omega_{\rm ti}$.  \citet{Voit_2017_BigPaper} showed that an internal gravity wave can resonate with pairs of sound waves having a beat frequency equal to the gravity-wave frequency,
and transfers wave energy at a rate of $\sim \Delta_r (kr)  \omega_{\rm buoy}$.  Saturation therefore occurs when $\Delta_r \sim (\omega_{\rm ti} / \omega_{\rm buoy})(kr)^{-1}$, implying a saturation amplitude for entropy perturbations of $| \delta_K | \sim \alpha_K (\omega_{\rm ti} / \omega_{\rm buoy})(kr)^{-1}$.  In the long-wavelength limit ($kr \sim 1$), this result accords with the one found in simulations.  \citet{Voit_2017_BigPaper} called this saturation mechanism ``buoyancy damping" because buoyant motions are what transfer kinetic energy away from the gravity wave.  The limiting amplitude that emerges does not depend on sound waves to be the dissipation channel.  Other dissipation mechanisms, such as drag and viscosity, may be present and can assist saturation as long as they can dissipate energy faster than it is being transferred, but they are not necessary for saturation.

\item  Further reduction in $\alpha_K$ fundamentally changes the nature of thermal instability, because it is no longer oscillatory for $\omega_{\rm buoy} / \omega_{\rm ti} \approx \alpha_K^{1/2} (t_{\rm cool} / t_{\rm ff}) \lesssim 1$.  Entropy perturbations then grow too rapidly for buoyancy to reduce the entropy contrasts responsible for local deviations from thermal balance.  The entropy contrast of a thermally unstable gas blob in this regime therefore increases monotonically.

\item  This relationship between local thermal instability and the large-scale entropy slope couples condensation with the global configuration of the system.   Condensation at the center of the system is inevitable because the saturation condition 
\begin{equation}
  \alpha_K^{1/2} (t_{\rm cool} / t_{\rm ff}) \gtrsim 1
\end{equation}
cannot be satisfied at the center of a hydrostatic medium unless the gas density is singular there.  Central thermal energy input owing to condensation-triggered feedback then boosts the global condensation rate because the convection it causes flattens the central entropy gradient and shuts off buoyancy damping.  If all of the feedback energy is centrally thermalized, then a condensation catastrophe can ensue until feedback raises the central cooling time of the ambient medium to $\gtrsim 1$~Gyr \citep[see, e.g.,][]{Meece_2017ApJ...841..133M}.

\item  Steady self-regulation therefore requires some of the feedback energy to be thermalized outside of the isentropic center, in a region with $\alpha_K^{1/2} (t_{\rm cool} / t_{\rm ff}) \gtrsim 1$ (i.e. the power-law zone).  Gas in an isothermal potential well consequently ends up reaching its minimum value of $t_{\rm cool} / t_{\rm ff}$ at the interface between the central isentropic zone and the power-law zone, if  $\alpha_K > 2/3$ in the power-law zone.   The magnitude of $t_{\rm cool} / t_{\rm ff}$ in the ambient gas at that interface is a measure of the system's susceptibility to condensation.

\end{enumerate}

This last point is where turbulence can enter the picture.  Linear thermal instability saturates before leading to condensation in an otherwise static ambient medium with $5 \lesssim \min (t_{\rm cool} / t_{\rm ff}) \lesssim 20$ and $\alpha_K \sim 2/3$, but moderate uplift velocities ($\sim v_c$) can stimulate condensation.  Uplift that is nearly adiabatic promotes condensation because it sharply increases $t_{\rm ff}$ while $t_{\rm cool}$ remains nearly constant in the uplifted gas.  If enough uplift occurs, local violations of the saturation condition arise and can permit condensation.  

Turbulence can potentially cause similar violations of the saturation condition by transporting low-entropy gas blobs to large enough altitudes.  Condensation can happen in such a medium if its turbulent velocity dispersion is comparable to the downward velocity that a condensing gas blob would ultimately develop as its buoyancy decreases.  One realization of this condensation mode is implicit in the simulation by \citet{Gaspari+2013MNRAS.432.3401G} with $\alpha_K^{1/2} (t_{\rm cool} / t_{\rm ff}) \approx 10$, in which forced turbulence with a velocity dispersion $\sigma_{\rm t} \approx 0.5 \sigma_v$ promotes inhomogeneous condensation. The following section develops these ideas further in order to illustrate how the critical value of $\min (t_{\rm cool} / t_{\rm ff})$ for condensation depends on the turbulent velocity dispersion.

\section{A Heuristic Model for Condensation}
\label{sec-Heuristic}

In pursuit of physical insight into the coupling between turbulence and condensation, this section explores the properties of a simple nonlinear dynamical model with just enough complexity to represent the stochastic momentum impulses characteristic of turbulence.  The model consists of two equations for the evolution of a perturbation's Eulerian entropy contrast $\delta_K$ and its dimensionless radial velocity $\Delta_v \equiv v_r / \sigma_v$.  Here, $v_r$ is the perturbation's radial velocity in a potential in which $\sigma_v = \, v_c \, / \sqrt{2} $ is constant with radius.  Buoyancy in an otherwise hydrostatic medium causes an acceleration $(2 \sigma_v / t_{\rm ff})(3 \delta_K / 5)$.  The perturbation's equations of motion are therefore
\begin{eqnarray}
  \dot{\Delta}_v &  = & \frac {6} {5 t_{\rm ff}} \delta_K - \omega_{\rm D} \Delta_v + \eta
    \label{eq-Delta_v_dot} \\
  \dot{\delta}_K &  =  & ( \omega_{\rm ti} - \omega_{\rm mix} ) \delta_K 
  			- \alpha_K \frac {\Delta_v} {t_{\rm ff}} 
	\label{eq-Delta_K_dot}
\end{eqnarray}
where $\omega_{\rm D}$ represents the buoyancy damping rate, $\omega_{\rm mix}$ represents the rate at which hydrodynamical mixing reduces the Eulerian entropy contrast, and $\eta$ is a source term for stochastic momentum impulses.  The perturbation's Langrangian entropy contrast and radial displacement are then determined by $\dot{\Delta}_K = (\omega_{\rm ti} - \omega_{\rm mix}) \delta_K$ and $\dot{\Delta}_r = \Delta_v / t_{\rm ff}$.  (See the Appendix for more details.)

Aside from the buoyancy term, Equation (\ref{eq-Delta_v_dot}) is a Langevin equation for the Brownian motion of a Lagrangian gas blob.  Such equations are used to model turbulent diffusion of particles in Earth's atmosphere \citep[e.g.,][]{Thomson_1984_Diffusion_Langevin} and are qualitatively suitable for modeling the diffusion of a circumgalactic gas blob that manages to remain coherent on timescales $\lesssim \omega_{\rm mix}^{-1}$.  If gravity is negligible, a random forcing function with the property $\langle \eta(t) \eta(t-\tau) \rangle = 2 \omega_{\rm D} (\sigma_{\rm t} / \sigma_v)^2 \delta(\tau)$, where angular brackets denote a time average and $\delta(\tau)$ is a Dirac delta function, produces a radial velocity dispersion $\langle \Delta_v^2 \rangle = (\sigma_{\rm t} /\sigma_v)^2$ that plays the role of turbulence in the model.

Combining the equations of motion to obtain a single noise-driven oscillator equation gives
\begin{equation}
  \ddot{\Delta}_v + (\omega_{\rm D} - \tilde{\omega}_{\rm ti} ) \dot{\Delta}_v
        + ( \tilde{\omega}_{\rm buoy}^2 - \omega_{\rm D} \tilde{\omega}_{\rm ti} ) \Delta_v = \tilde{\eta}
       \; \; ,
\end{equation}
where $\tilde{\omega}_{\rm ti}  \equiv  \omega_{\rm ti} - \omega_{\rm mix} - \dot{\Delta}_r$, $\tilde{\omega}_{\rm buoy}^2 \equiv \omega_{\rm buoy}^2 + \dot{\omega}_{\rm D}$, and $\tilde{\eta}  \equiv \dot{\eta} - \tilde{\omega}_{\rm ti} \eta$.  Plugging in a disturbance in which all perturbed quantities are $\propto e^{- i \omega t}$ then yields the dispersion relation
\begin{equation}
  \omega^2 + i (\omega_{\rm D} - \tilde{\omega}_{\rm ti}) \omega 
    - ( \tilde{\omega}_{\rm buoy}^2 - \omega_{\rm D} \tilde{\omega}_{\rm ti} ) = 0
\end{equation}
in the absence of noise.  The frequency solutions are
\begin{equation}
  \omega_\pm = i \frac {\tilde{\omega}_{\rm ti} - \omega_{\rm D}} {2}
       \pm \tilde{\omega}_{\rm buoy} 
       		\left[ 1 - \frac {(\tilde{\omega}_{\rm ti} + \omega_{\rm D})^2} {4 \tilde{\omega}_{\rm buoy}^2}
		\right]^{1/2}
	\; \; ,
\end{equation}
and they lead to thermal instability for $\tilde{\omega}_{\rm ti} > \omega_{\rm D}$ and to overdamping for $\omega_{\rm D} > 2 \omega_{\rm buoy} - \tilde{\omega}_{\rm ti}$.

\subsection{The Linear Regime}

\begin{figure*}[t]
\begin{center}
\includegraphics[width=5.2in, trim = 0.1in 0.2in 0.0in 0.0in]{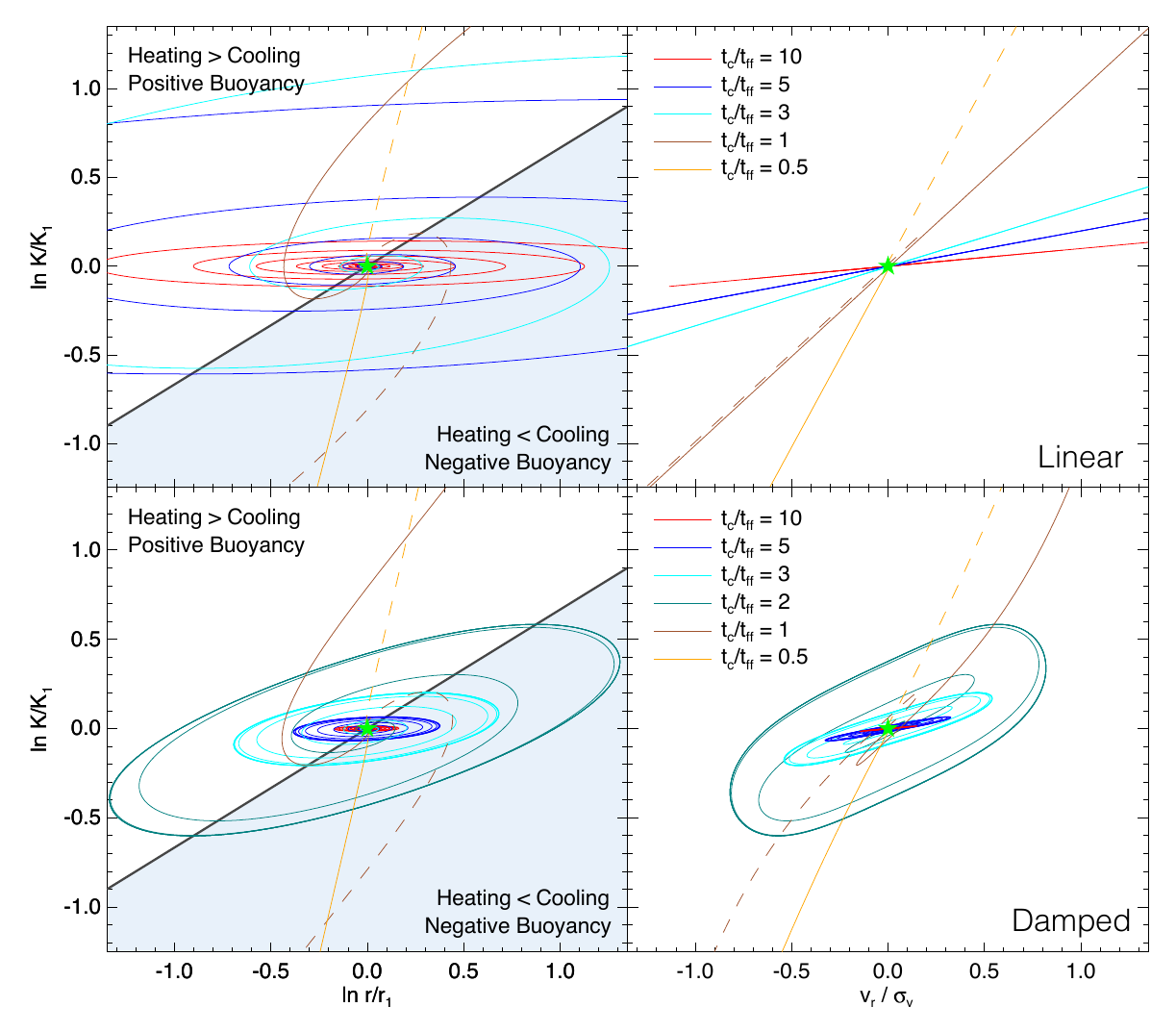} \\
\end{center}
\caption{ \footnotesize 
Trajectories of thermally unstable internal gravity waves in the $r$--$K$ plane (left) and the $v_r$--$K$ plane (right).  Top panels show linearized trajectories that have been extrapolated to nonlinear wave amplitudes.  Bottom panels show how those trajectories change when buoyancy damping is included.  A thick charcoal line in the $r$--$K$ plane shows the background entropy profile, $\bar{K}(r) = K_1 (r/r_1)^{2/3}$, for which $t_{\rm cool} / t_{\rm ff}$ is constant in an isothermal potential.  Below that line, in the shaded regions, a perturbation in pressure equilibrium is denser than its surroundings, which leads to net cooling and downward acceleration.  Above that line, a perturbation is less dense than its surroundings, gains heat energy, and accelerates upward.  Gravity waves are therefore thermally unstable, with amplitudes that grow on a timescale $\sim t_{\rm cool}$.  Colored lines show how the evolution of the amplitudes depends on $t_{\rm cool} / t_{\rm ff}$, with red lines representing $t_{\rm cool} / t_{\rm ff} = 10$, blue lines representing $t_{\rm cool} / t_{\rm ff} = 5$, cyan lines representing $t_{\rm cool} / t_{\rm ff} = 3$, teal lines representing $t_{\rm cool} / t_{\rm ff} = 2$, brown lines representing $t_{\rm cool} / t_{\rm ff} = 1$, and orange lines representing $t_{\rm cool} / t_{\rm ff} = 0.5$.  Solid lines start with $v_r = 0.01 \sigma_v$ and dashed ones start with $v_r = - 0.01 \sigma_v$.  A green star marks the starting point.  Trajectories in the bottom panels converge to limit cycles because energy transfer into dissipative sound waves increases as the amplitude of a thermally unstable gravity wave grows.  Thermal instability in media with $t_{\rm cool} / t_{\rm ff} \gg 1$ therefore saturates when $\omega_{\rm D} = \omega_{\rm ti}$, at which point the rms amplitudes are $\langle \Delta_r^2 \rangle^{1/2} \approx \langle \Delta_v^2 \rangle^{1/2} \approx t_{\rm ff} /t_{\rm cool}$ and $\langle \Delta_K^2 \rangle^{1/2} \approx (t_{\rm ff} /t_{\rm cool})^2$   
\vspace*{1em}
\label{fig-Linear_Damped}}
\end{figure*}

For sufficiently small oscillations, both damping and mixing can be ignored and $t_{\rm ff}$ remains effectively constant.  The dispersion relation then reduces to
\begin{equation}
  \omega^2 - i \omega_{\rm ti} \omega - \omega_{\rm buoy}^2 = 0
  \; \; .
\end{equation}
It is essentially identical to Equation (\ref{eq-gmodes}) for thermally unstable internal gravity waves with wavevectors oriented perpendicular to gravity.  The top panels of Figure~\ref{fig-Linear_Damped} show how the corresponding disturbances evolve for different values of $t_{\rm cool} / t_{\rm ff}$.  An entropy slope $\alpha_K = 2/3$ is assumed so that $t_{\rm cool} / t_{\rm ff}$ remains constant with radius.

\subsection{Oscillations with Buoyancy Damping}

According to the analysis of \citet{Voit_2017_BigPaper}, internal gravity waves in the circumgalactic medium transfer their kinetic energy to dissipative sound waves at a rate $\omega_{\rm D} \approx | \Delta_v | \, (kr) \, \omega_{\rm buoy}$.  The dispersion relation becomes
\begin{equation}
    \omega^2 + i ( \omega_{\rm D} - \omega_{\rm ti}) \omega - \omega_{\rm buoy}^2 = 0  \; \; ,
\end{equation}
as long as changes in the dissipation rate are slow compared to the frequency of buoyant oscillations. 
Long-wavelength ($kr \approx 1$) gravity waves of linear amplitude therefore grow via thermal instability until $\omega_{\rm D} = \omega_{\rm ti}$, at which point $| \Delta_v | \approx  (\omega_{\rm ti} / \omega_{\rm buoy})$.  The asymptotic trajectory is a limit cycle toward which the amplitudes of both larger and smaller disturbances converge.  The bottom panels of Figure~\ref{fig-Linear_Damped} illustrate some examples based on setting $\omega_{\rm D} = |\Delta_v| \omega_{\rm buoy}$. 

Disturbances of sufficiently small wavelength can be overdamped, in which case condensation is monotonic at a rate $\sim \omega_{\rm ti}$ as long as mixing is suppressed \citep[e.g.,][]{Nulsen_1986MNRAS.221..377N}.  However, thermally unstable gravity waves growing from the linear regime in a system with $\omega_{\rm ti} < \omega_{\rm buoy}$ reach the limit cycle before becoming overdamped.  In order to become overdamped, a gravity wave must be driven by external forcing past the saturation limit at $| \Delta_v | \approx (\omega_{\rm ti} / \omega_{\rm buoy}) (kr)^{-1}$ to a velocity amplitude $| \Delta_v | \gtrsim [ 2 - (\omega_{\rm ti} / \omega_{\rm buoy})] (kr)^{-1}$.  Those perturbations then have the opportunity to condense without experiencing any more oscillations, but if their wavelengths are short, they are likely to be rather fragile and easily shredded by hydrodynamic mixing.  Furthermore, short-wavelength perturbations are unlikely to fuel a well-coupled feedback response after they condense because they have small terminal velocities and descend slowly toward the bottom of the potential well, where the accretion engine lies.  The rest of the discussion therefore focuses primarily on condensation of long-wavelength perturbations.

\subsection{Mixing}

The hydrodynamic mixing rate is less certain than the buoyancy damping rate, but thermal instability obviously cannot grow if $\omega_{\rm mix} > \omega_{\rm ti}$ \cite[for more details, see][and the Appendix]{BanerjeeSharma_2014MNRAS.443..687B,Gaspari_2017MNRAS.466..677G,Gaspari_2018ApJ...854..167G}.  Hydrodynamical simulations that model the shredding of a dense gas blob moving through a low-viscosity medium typically find that the blob is effectively destroyed after it has passed through a column density of ambient gas equivalent to several times its own column density \citep[e.g.,][]{KleinColellaMcKee_1990ASPC...12..117K,StoneNorman_1992ApJ...390L..17S,KleinMcKeeColella_1994ApJ...420..213K}.  The ratio $\omega_{\rm D} / \omega_{\rm mix}$ can therefore be at least several times greater than unity, and possibly larger.  For example, simulations by \citet{McCourt_2015MNRAS.449....2M} have shown that magnetic fields can preserve a condensing blob against mixing for a much longer time period, which promotes condensation.  

Parameterizing the uncertainties surrounding mixing by defining $f_{\rm mix} \equiv \omega_{\rm mix} / \omega_{\rm D}$ enables an assessment of its role.  Thermal instability can proceed faster than mixing as long as
\begin{equation}
      f_{\rm mix}  \lesssim \frac {\omega_{\rm ti}} {| \Delta_v | (kr) \, \omega_{\rm buoy}} 
         \; \; ,
\end{equation}
but a perturbation driven into the overdamped regime has $| \Delta_v | (kr) \gtrsim 2$.  Stimulating condensation by driving overdamped gravity waves therefore requires $f_{\rm mix} \lesssim \omega_{\rm ti} / 2 \omega_{\rm buoy} \approx (4 \alpha_K)^{-1/2} (t_{\rm ff} / t_{\rm cool})$.

This feature of the model offers an opportunity to tune the $t_{\rm ff} / t_{\rm cool}$ threshold for condensation simply by choosing a particular value for $f_{\rm mix}$.  Instead of taking advantage of that opportunity, most of this paper's calculations set $\omega_{\rm mix}$ to zero and gauge a perturbation's susceptibility to mixing by tracking the quantity
\begin{equation}
  L_{\rm path}(t) \equiv \sigma_v \int_0^t |\Delta_v| \, dt
  \; \; ,
\end{equation}
which is a measure of the total amount of shear to which the perturbation has been exposed.  The product $k L_{\rm path}$ is approximately equal to the column density of the ambient medium along the perturbation's path divided by the column density of the perturbation itself. 

\subsection{Nonlinearity of the Restoring Force}

\begin{figure*}[t]
\begin{center}
\includegraphics[width=5.2in, trim = 0.0in 0.2in 0.0in 0.0in]{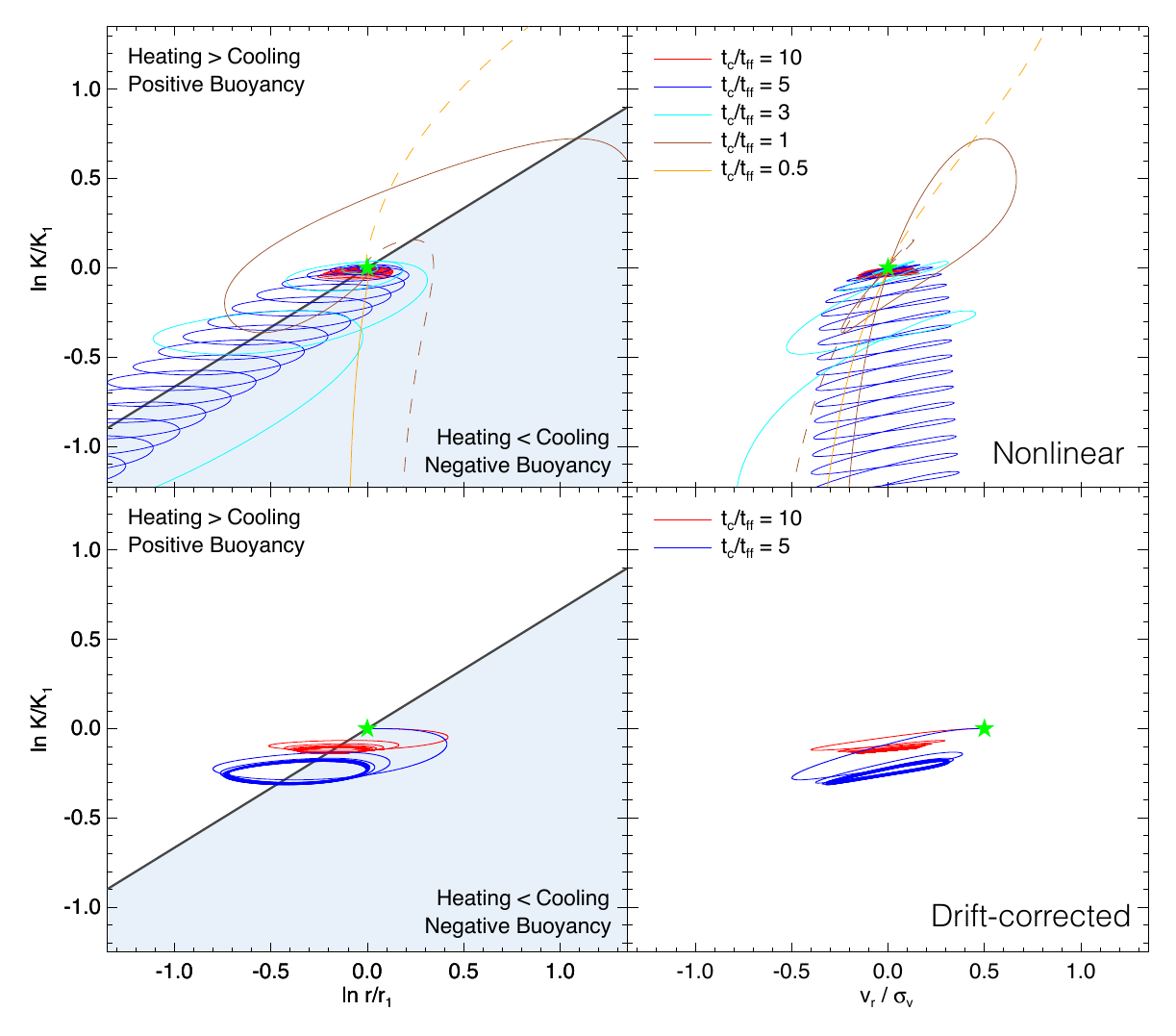} \\
\end{center}
\caption{ \footnotesize 
Centroid drift of thermally unstable perturbations with $kr = 1$ when nonlinearity of the buoyant restoring force is included.  Lines and panel axes are as in Figure~\ref{fig-Linear_Damped}.  The top panels show how variations in the local value of $t_{\rm ff}$ cause a perturbation to drift toward lower entropy. Oscillating disturbances spend more time on the net cooling (shaded) side of the mean entropy profile than on the net heating (unshaded) side.  As a result, each successive cycle lowers the time-averaged entropy of the perturbation.  That drift can be corrected using Equation (\ref{eq-DriftCorrection}), which gives the trajectories shown in the bottom panels.  The drift correction mimics numerical simulations in which the average heating rate on each equipotential surface is forced to match the average cooling rate on that surface.  Each trajectory in the bottom panels starts with $v_1 = 0.5 \, \sigma_v$ so that it approaches its limit cycle from the high-amplitude side.  The result of an outward velocity impulse in such a balanced system is an initial decrease in the perturbation's entropy, followed by limit cycle oscillations around a new centroid.
\vspace*{1em}
\label{fig-Nonlinear_Nodrift}}
\end{figure*}

Radial motion introduces more nonlinearity into the gravity wave oscillator by periodically altering the dynamical timescale.  The clock speed at which at which dynamical processes unfold depends on the local value of the freefall time, which is $t_{\rm ff} \propto r$ in this realization.  A perturbation consequently tends to accelerate more slowly at higher altitudes than at lower ones \citep[see][]{Kepler_1609} and therefore spends more time in a net cooling state than in a net heating state.  The top panels in Figure~\ref{fig-Nonlinear_Nodrift} show the drift in $\langle \Delta_K \rangle$ that results.  

In order to compute the drift, non-linear corrections need to be made to both $t_{\rm ff}$ and $\omega_{\rm ti}$.  The freefall time correction is relatively simple:
\begin{equation}
  t_{\rm ff} (\Delta_r) = t_{\rm ff} (r_1) \, \frac {r} {r_1} = t_{\rm ff} (r_1) \, e^{\Delta_r}
  \label{eq-tff}
  \; \; .
\end{equation}
The correction to $\omega_{\rm ti}$ depends on the cooling function $\Lambda (T)$ and can be expressed as
\begin{equation}
   \omega_{\rm ti} (\Delta_r,\delta_K) = 
       \omega_{\rm ti} (r_1)  \, e^{-\Delta_r} e^{- \alpha_{\rm ti} \delta_K}
       \label{eq-omega_ti}
\end{equation}
in a medium in which the background value of $t_{\rm cool} / t_{\rm ff}$ is constant with radius.
The $e^{-\Delta_r}$ factor accounts for the $\omega_{\rm ti}(r)/\omega_{\rm ti}(r_1)$ ratio in an ambient medium with a constant $t_{\rm cool}/t_{\rm ff}$ ratio, while the $e^{- \alpha_{\rm ti} \delta_K}$ factor accounts for the cooling-time contrast between the perturbation and the local  ambient medium.  In pressure equilibrium, the cooling-time contrast is $\propto K^{(6 - 3 \lambda)/5}$ in a medium with $\Lambda \propto T^\lambda$.  Setting $\alpha_{\rm ti} = (6 - 3 \lambda)/5$ therefore provides the necessary dependence of $\omega_{\rm ti}$ on entropy contrast.
The following calculations set $\lambda = 0$, which is appropriate for circumgalactic gas at $10^7$~K with approximately solar abundances.  Hotter gas has $\lambda \approx 0.5$, and cooler gas (down to $\sim 10^5$~K) has $\lambda \approx -1$, but those differences do not change the qualitative results.

\subsection{Maintaining Thermal Balance}

Time-averaged thermal balance can be restored to the perturbed system by adding a constant heating term that offsets the centroid drift, but a single heating correction cannot simultaneously restore time-averaged thermal balance to all perturbations.  For example, drifts such as those shown in the top panels of Figure \ref{fig-Nonlinear_Nodrift} can be largely offset by adding a constant heating term
\begin{equation}
  \langle \dot{\Delta}_K \rangle_{\rm corr} = 0.83 \, \omega_{\rm ti} 
  				\left( \frac {t_{\rm ff}} {t_{\rm cool}} \right)^2 (kr)^{-2}
	\label{eq-DriftCorrection}
\end{equation}
to the right-hand side of Equation (\ref{eq-Delta_K_dot}).  This correction produces the trajectories shown in the bottom of Figure \ref{fig-Nonlinear_Nodrift} for perturbations with $kr = 1$.   It removes almost all of the drift from those perturbations in environments with $t_{\rm cool} / t_{\rm ff} \gtrsim 4$ but depends strongly on the perturbation's wavelength.  The required correction is smaller for short-wavelength perturbations than for long-wavelength ones because the short-wavelength perturbations reach a limit cycle at smaller amplitudes and are subject to less drift.

In numerical simulations of thermal instability the average heating rate in each equipotential layer can be forced to balance the average cooling rate of the layer \citep[e.g.,][]{McCourt+2012MNRAS.419.3319M,Gaspari+2013MNRAS.432.3401G,Meece_2015ApJ...808...43M,ChoudhurySharma_2016MNRAS.457.2554C}.  A given uniform heating rate can balance perturbations on the limit cycle at one particular wavelength but falls short of the time-averaged cooling of longer-wavelength perturbations and overheats perturbations of shorter wavelength.  The heating rate needed to keep the overall layer in thermal balance therefore depends on the equilibrium power spectrum of entropy perturbations, which depends in turn on mixing and other uncertain features of the model.  Given those uncertainties, this paper simply applies a drift correction equivalent to Equation (\ref{eq-DriftCorrection}) with $kr = 1$, because most of the excess cooling arising from asymmetric oscillations comes from the long-wavelength perturbations.

\subsection{Episodic Heating}

In the circumgalactic media of real galaxies, global thermal balance is approximate, not exact, and may require episodic heating modulated by fluctuations in condensation.  This paper will not consider episodic heating in detail, in part because its qualitative effects on condensation are secondary to those of buoyancy, as long as the heating mechanism satisfies three important conditions:
\begin{enumerate}
\item Heat input needs to maintain a positive entropy slope large enough to ensure $\alpha_K^{1/2} (t_{\rm cool} / t_{\rm ff}) \gtrsim 1$. 
\item The time intervals between heating episodes cannot be longer than the local cooling time.
\item The net heating rate, integrated over time, must be positive for $\delta_K > 0$ and negative for $\delta_K < 0$.  
\end{enumerate}
If condition 1 is violated, then convection eliminates the entropy gradient needed for buoyancy damping and a cooling catastrophe results \citep[e.g.,][]{Meece_2017ApJ...841..133M}.  If condition 2 is violated, then a catastrophic cooling flow ensues, in which thermal instability does not produce a multiphase medium until $t_{\rm cool} \lesssim t_{\rm ff}$ \citep[e.g.,][]{LiBryan2012ApJ...747...26L}.  If condition 3 is violated, then $\langle \omega_{\rm ti} \rangle < 0$, and thermal instability cannot proceed.  However, it is reasonable to assume that all three conditions usually hold in the circumgalactic media around sufficiently massive galaxies \citep{Voit_2017_BigPaper}.

\subsection{Transition from Buoyant to Ballistic Motion}

\begin{figure*}[t]
\begin{center}
\includegraphics[width=4.5in, trim = 0.0in 0.2in 0.0in 0.0in]{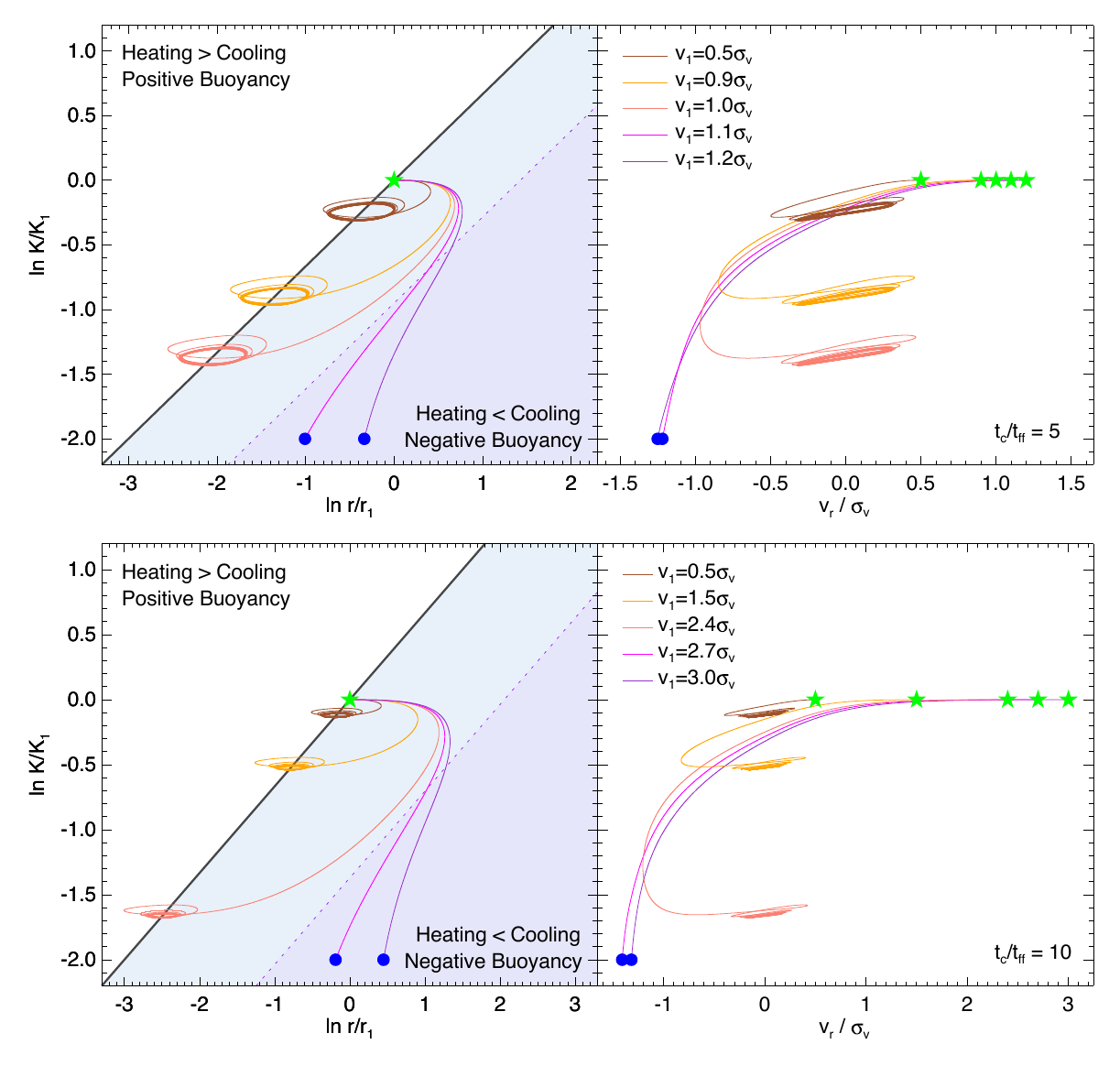} \\
\end{center}
\caption{ \footnotesize 
Trajectories of perturbations that start with differing initial radial speeds ($v_1$) in environments with $t_{\rm cool} / t_{\rm ff} = 5$ (top panels) and $t_{\rm cool} / t_{\rm ff} = 10$ (bottom panels).  Panel axes are as in Figure \ref{fig-Linear_Damped}.  The entropy slope is $\alpha_K = 2/3$, so that $t_{\rm cool} / t_{\rm ff} $ is constant with radius, and the thick charcoal line indicates the background entropy profile.  A dotted purple line in each of the left panels shows the condensation criterion from Equation (\ref{eq-ApproxCondensationCondition}), which is satisfied in the lavender region below the line.  Colored lines in the top panels show trajectories with $v_1$ equal to $0.5 \sigma_v$ (brown line), $0.9 \sigma_v$ (orange line), $1.0 \sigma_v$ (salmon line), $1.1 \sigma_v$ (magenta line), and $1.2 \sigma_v$ (violet line).  Colored lines in the bottom panels show trajectories with $v_1$ equal to $0.5 \sigma_v$ (brown line), $1.5 \sigma_v$ (orange line), $2.4 \sigma_v$ (salmon line), $2.7 \sigma_v$ (magenta line), and $3.0 \sigma_v$ (violet line). Green stars show where the trajectories begin, and blue circles indicate trajectories ending in condensation.
\vspace*{1em}
\label{fig-Ballistic}}
\end{figure*}

Condensation progresses as long as radiative cooling proceeds more rapidly than buoyancy damping.  That happens while $\dot{\delta}_K < 0$, which is equivalent to 
\begin{equation} 
  \delta_K < - \frac {1} {\alpha_{\rm ti}} 
  		\ln \left[ \frac {\alpha_K} {\alpha_{\rm ti}} 
		             \left| \frac {\Delta_v} {\delta_K} \right|
		             \left( \frac {t_{\rm cool}} {t_{\rm ff}} \right)
		             \right]
	\label{eq-CondensationCondition}
\end{equation}
when mixing is negligible.  In this expression, the ratio $t_{\rm cool} / t_{\rm ff}$ represents the background environment and not the ratio within the perturbation.  As a low-entropy perturbation accelerates downward, thereby raising $\left| \Delta_v \right|$, the condition becomes increasingly restrictive until the descending gas approaches its terminal velocity.  If the condition in Equation (\ref{eq-CondensationCondition}) is still satisfied at that point, then condensation is assured, and the resulting dense gas blob subsequently follows a ballistic trajectory modified by hydrodynamic drag.

A precise calculation of the condensing blob's terminal velocity is well beyond the scope of this simple model, because it cannot account for deformations of the perturbation as it is compressed and shredded by its motion through the ambient medium.  However, an estimate is possible if the column density of the perturbation along its direction of motion remains approximately constant.  If that is the case, then $\omega_{\rm D} \approx | \Delta_v | (kr) t_{\rm ff}^{-1}$.  In making the estimate, one also needs to recognize that Equation (\ref{eq-Delta_v_dot}) must be modified when $\delta_K  < - 5/3$, because otherwise the term representing acceleration due to negative buoyancy would exceed the free gravitational acceleration.  The calculations presented in this paper account for this circumstance by substituting
\begin{equation}
    \dot{\Delta}_v = \frac {2} {t_{\rm ff}} 
         \left[  \frac {\delta_K} { \max \left( \frac {5} {3} ,  | \delta_K | \right) } \right]
    	- \omega_{\rm D} \Delta_v + \eta
    \label{eq-Delta_v_dot_mod} 
\end{equation}
for Equation (\ref{eq-Delta_v_dot}).  With this crude modification, the asymptotic speeds of both low-entropy condensates and high-entropy bubbles are properly limited, even though the computed trajectories are not exact.  

The appropriate value for $\Delta_v / \delta_K$ in Equation (\ref{eq-CondensationCondition}) can then be determined by setting $\dot{\Delta}_v = 0$ and $\eta = 0$ in Equation (\ref{eq-Delta_v_dot_mod}):
\begin{equation}
  \frac {\Delta_v} {\delta_K} \approx \left( \frac {2} {kr} \right)^{1/2} 
  		\left[  \max \left( \frac {5} {3} | \delta_K | , \delta_K^2 \right) \right]^{-1/2}
		\label{eq-Deltarat}
		\; \; .
\end{equation}
Together, Equations (\ref{eq-CondensationCondition}) and (\ref{eq-Deltarat}) yield an approximate relation between the $t_{\rm cool} / t_{\rm ff}$ ratio of a thermally balanced galactic atmosphere and the critical Eulerian entropy contrast $\delta_{\rm c}$ that a gravity-wave perturbation needs to achieve in order to ensure condensation:
\begin{equation}
  \frac {t_{\rm cool}} {t_{\rm ff}} \approx \frac {\alpha_{\rm ti}} {\alpha_K}
  		\left[ \frac {kr} {2} | \delta_{\rm c} | 
		         \max \left( \frac {5} {3} , | \delta_{\rm c} | \right) \right]^{1/2}
		    e^{- \alpha_{\rm ti} \delta_{\rm c}}
		\; \; .
  	\label{eq-ApproxCondensationCondition}
\end{equation}
(See the Appendix for a discussion of the relationship between this criterion, in which $t_{\rm cool}/t_{\rm ff}$ represents the ambient medium, and criteria based on the value of $t_{\rm cool}/t_{\rm ff}$ local to the perturbation.)  
Some representative values of the critical contrast for long-wavelength perturbations ($kr = 1$) in a medium with $\alpha_K = 2/3$ and $\alpha_{\rm ti} = 6/5$ are $\delta_{\rm c}(t_{\rm cool} / t_{\rm ff} = 5) = -0.95$, $\delta_{\rm c}(t_{\rm cool} / t_{\rm ff} = 10) = -1.37$, and $\delta_{\rm c}(t_{\rm cool} / t_{\rm ff} = 20) = -1.80$.  Such perturbations transition to drag-limited ballistic motion when $\Delta_K - \alpha_K \Delta_r < \delta_{\rm c}$ and subsequently condense.

\begin{figure*}[t]
\begin{center}
\includegraphics[width=5.0in, trim = 0.0in 0.0in 0.0in 0.0in]{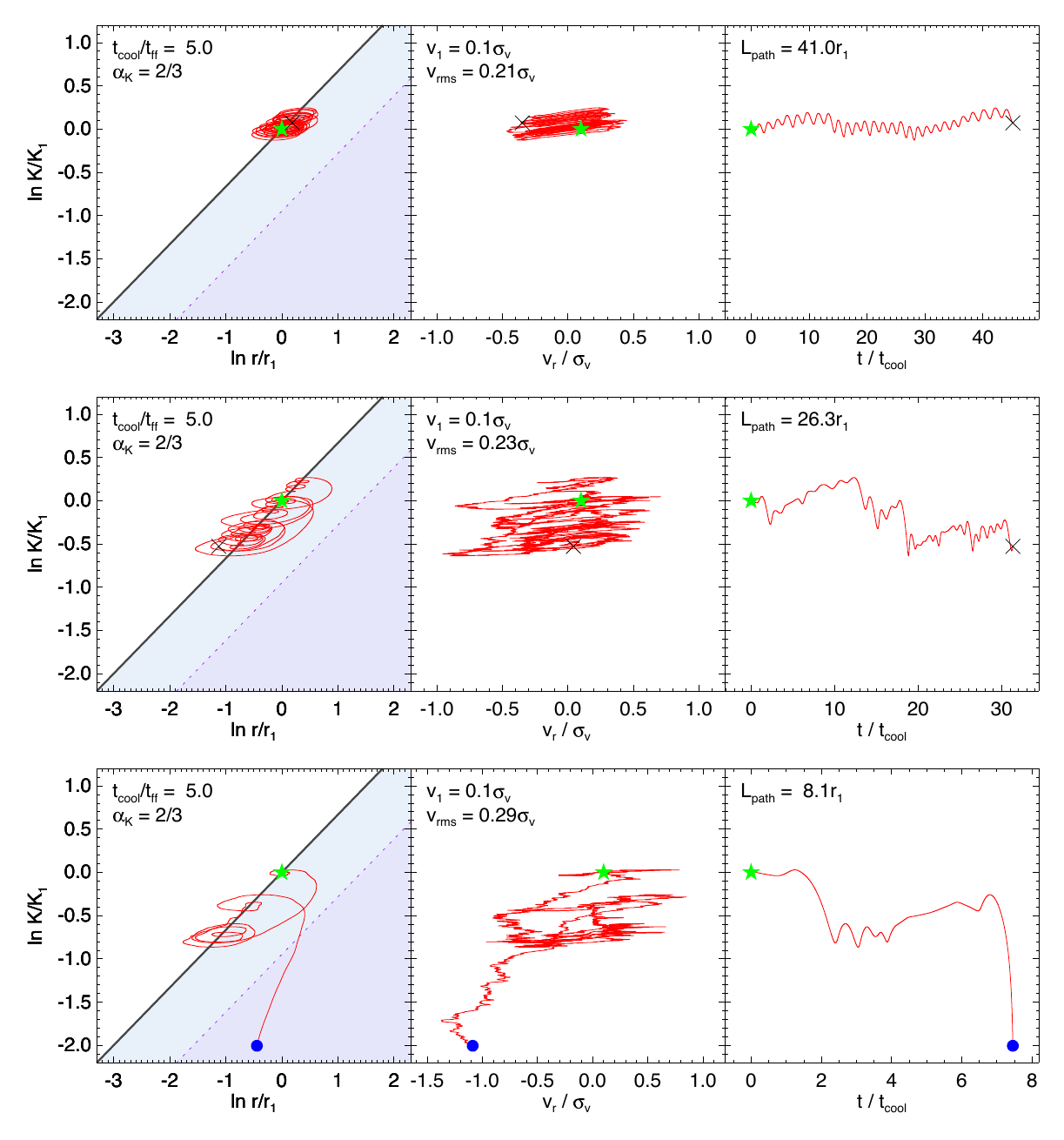} \\
\end{center}
\caption{ \footnotesize 
Trajectories of long-wavelength ($kr=1$) gravity wave perturbations driven by noise in a medium with $t_{\rm cool} / t_{\rm ff} = 5$ and $\alpha_K = 2/3$.  Each row of panels shows a single trajectory in the $r$--$K$ plane (left), in the $v_r$--$K$ plane (middle), and as a function of time in units of the initial cooling time (right).  Shaded regions in the left panels show where cooling exceeds heating, and perturbations that cross the dotted line into the lavender region are destined to condense.  For each trajectory, the random sequence of momentum impulses that drives the oscillator is the same, except for the gain factor $\eta_0$.  In the top set of panels, a small gain factor leads to some diffusion around the limit cycle and produces an rms velocity dispersion $v_{\rm rms} = 0.21 \sigma_v$.  A moderate gain factor produces more diffusion and a larger velocity dispersion, as shown in the middle set of panels.  Further increases in the gain factor ultimately force the oscillator into overdamping and condensation, as shown in the bottom set of panels.  In that case the velocity dispersion of the condensing trajectory is $0.29 \sigma_v$, and a blue endpoint signifies condensation.  Trajectories that do not condense end with a black cross.
\vspace*{1em}
\label{fig-blob_orbits_5}}
\end{figure*}

Figure \ref{fig-Ballistic} shows relationships between the criterion in Equation (\ref{eq-ApproxCondensationCondition}) and perturbation trajectories computed from the oscillator model for different initial values of $v_1$ and zero noise.  Perturbations that start with small outward velocities initially drift to lower entropy but eventually return to the mean profile and converge to limit cycles.  However, perturbations with sufficiently large initial velocities eventually satisfy the condensation criterion and approach terminal velocity while continuing to lose entropy.  The two different outcomes bifurcate around the line defined by Equation (\ref{eq-ApproxCondensationCondition}).

\begin{figure*}[t]
\begin{center}
\includegraphics[width=5.0in, trim = 0.0in 0.0in 0.0in 0.0in]{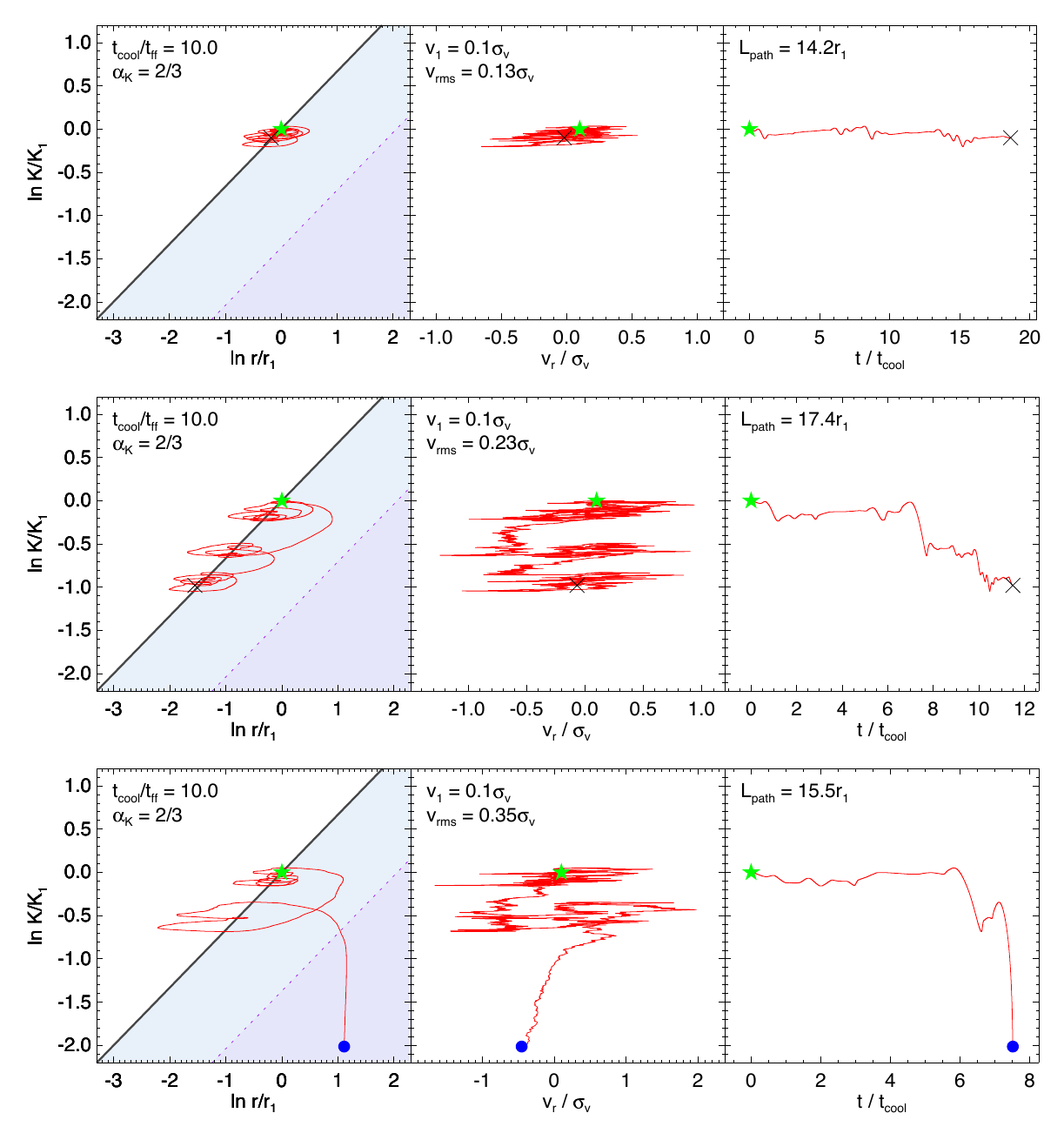} \\
\end{center}
\caption{ \footnotesize
Trajectories of long-wavelength ($kr=1$) gravity wave perturbations driven by noise in a medium with $t_{\rm cool} / t_{\rm ff} = 10$ and $\alpha_K = 2/3$.  Each row of panels shows a single trajectory in the $r$--$K$ plane (left), in the $v_r$--$K$ plane (middle), and as a function of time in units of the initial cooling time (right).  Shaded regions in the left panels show where cooling exceeds heating, and perturbations that cross the dotted line into the lavender region are destined to condense.  For each trajectory, the random sequence of momentum impulses that drives the oscillator is the same, except for the gain factor $\eta_0$.  In the top set of panels, a small gain factor leads to some diffusion around the limit cycle and produces an rms velocity dispersion $v_{\rm rms} = 0.13 \sigma_v$.  A moderate gain factor produces more diffusion and a larger velocity dispersion, as shown in the middle set of panels.  Further increases in the gain factor ultimately force the oscillator into overdamping and condensation, as shown in the bottom set of panels.  In that case the velocity dispersion of the condensing trajectory is $0.35 \sigma_v$, and a blue endpoint signifies condensation.  Trajectories that do not condense end with a black cross.
\vspace*{1em}
\label{fig-blob_orbits_10}}
\end{figure*}

Notice that the initial outward velocities required to produce condensation are hard to achieve in systems with $t_{\rm cool} / t_{\rm ff} \gtrsim 10$.  The bottom panels in Figure~\ref{fig-Ballistic} show that perturbations with initial velocities $v_1 \approx 2.5 \sigma_v$ are necessary to produce condensation in systems with $t_{\rm cool} / t_{\rm ff} = 10$.  For comparison, the ambient sound speed of hydrostatic gas with $\alpha_K = 2/3$ in an isothermal potential is $c_s \approx 1.8 \sigma_v$.  The required initial speeds are large because motion through the ambient medium saps the perturbation's initial kinetic energy as it attempts to rise to $\Delta_r = \ln (r/r_1) \approx 1$.

\begin{figure*}[t]
\begin{center}
\includegraphics[width=5.0in, trim = 0.0in 0.0in 0.0in 0.0in]{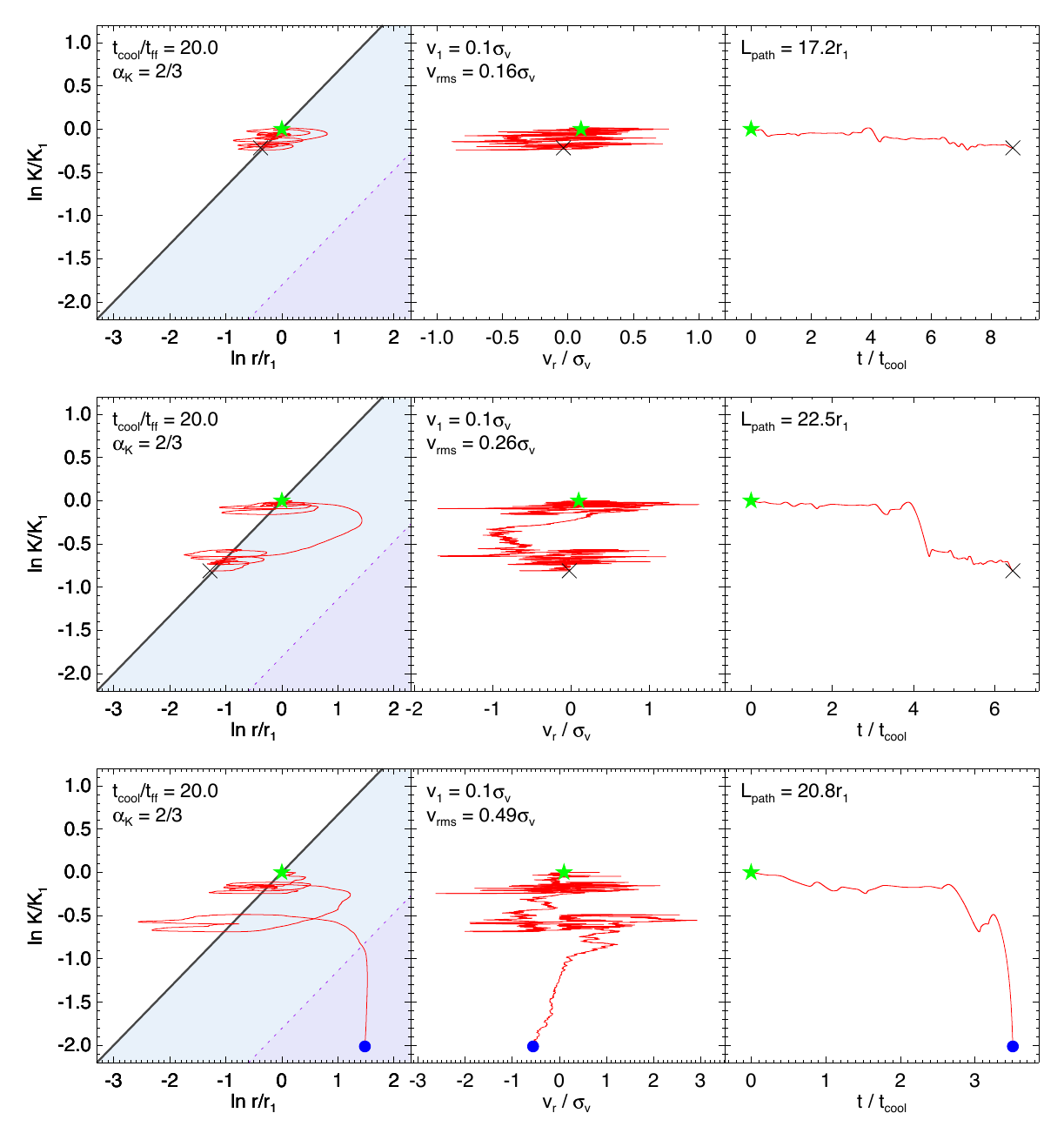} \\
\end{center}
\caption{ \footnotesize 
Trajectories of long-wavelength ($kr=1$) gravity wave perturbations driven by noise in a medium with $t_{\rm cool} / t_{\rm ff} = 20$ and $\alpha_K = 2/3$.  Each row of panels shows a single trajectory in the $r$--$K$ plane (left), in the $v_r$--$K$ plane (middle), and as a function of time in units of the initial cooling time (right).  Shaded regions in the left panels show where cooling exceeds heating, and perturbations that cross the dotted line into the lavender region are destined to condense.  For each trajectory, the random sequence of momentum impulses that drives the oscillator is the same, except for the gain factor $\eta_0$.  In the top set of panels, a small gain factor leads to some diffusion around the limit cycle and produces an rms velocity dispersion $v_{\rm rms} = 0.15 \sigma_v$.  A moderate gain factor produces more diffusion and a larger velocity dispersion, as shown in the middle set of panels.  Further increases in the gain factor ultimately force the oscillator into overdamping and condensation, as shown in the bottom set of panels.  In that case the velocity dispersion of the condensing trajectory is $0.49 \sigma_v$, and a blue endpoint signifies condensation.  Trajectories that do not condense end with a black cross.
\vspace*{1em}
\label{fig-blob_orbits_20}}
\end{figure*}

Inducing condensation is not quite as difficult in a bulk outflow that moves ballistically without hydrodynamic resistance.  \citet{Voit_2017_BigPaper} showed that such an outflow in a medium with $t_{\rm cool} / t_{\rm ff} \approx 10$ can produce condensation as long as $v_1 \gtrsim  2 \sigma_v$. The required outflow velocity is significantly larger than the rms velocity dispersions of molecular clouds typically observed in galaxy cluster cores \citep{Russell_2016MNRAS.458.3134R}.  However, there are systems in which up to $\sim 20$\% of the molecular gas may be moving outward at the required speeds \citep{McNamara_2014ApJ...785...44M}

\subsection{Turbulent Forcing of Condensation}

The main point of this paper is that turbulence with a 1D velocity dispersion $\sigma_{\rm t} \approx 0.5 \sigma_v$ can induce condensation in circumgalactic media with $\alpha_K \approx 2/3$ and $10 \lesssim t_{\rm cool} / t_{\rm ff} \lesssim 20$, without the need for large outflow velocities.  Figures \ref{fig-blob_orbits_5} through \ref{fig-blob_orbits_20} show how gravity wave oscillations progress into condensation as they are driven with increasing amounts of noise.  Each figure shows three trajectories of long-wavelength ($kr=1$) perturbations calculated by jointly integrating Equations (\ref{eq-Delta_v_dot_mod}) and (\ref{eq-Delta_K_dot}) while including the drift correction given by Equation (\ref{eq-DriftCorrection}).  The dissipation rate is set to $\omega_{\rm D} = | \Delta_v | (kr) \omega_{\rm buoy}$, Equation (\ref{eq-tff}) gives $t_{\rm ff}$, Equation (\ref{eq-omega_ti}) gives $\omega_{\rm ti}$, and $f_{\rm mix}$ is set to zero.  Noise is added with a sequence of $\eta$ values that are Markovian and Gaussian with $\langle \eta \rangle = 0$ and $\langle \eta^2 \rangle = \eta_0^2$.   Manipulating the gain factor $\eta_0$ adjusts the noise level.

Within each of the figures the only model parameter that changes is $\eta_0$.  The momentum impulses applied to the perturbation are otherwise identical.  The top row of panels in each figure illustrates a trajectory with only a small amount of noise that causes some diffusion around the limit cycle.  In the middle row is a trajectory with a larger noise amplitude that causes more diffusion, a larger velocity dispersion, and increased drift to lower entropy values.  Most of the drift happens during the biggest excursions to larger radii.  The bottom row of each figure shows a trajectory with an even larger noise amplitude that is sufficient to cause condensation.  

\begin{figure*}[t]
\begin{center}
\includegraphics[width=6.2in, trim = 0.0in 0.2in 0.0in 0.0in]{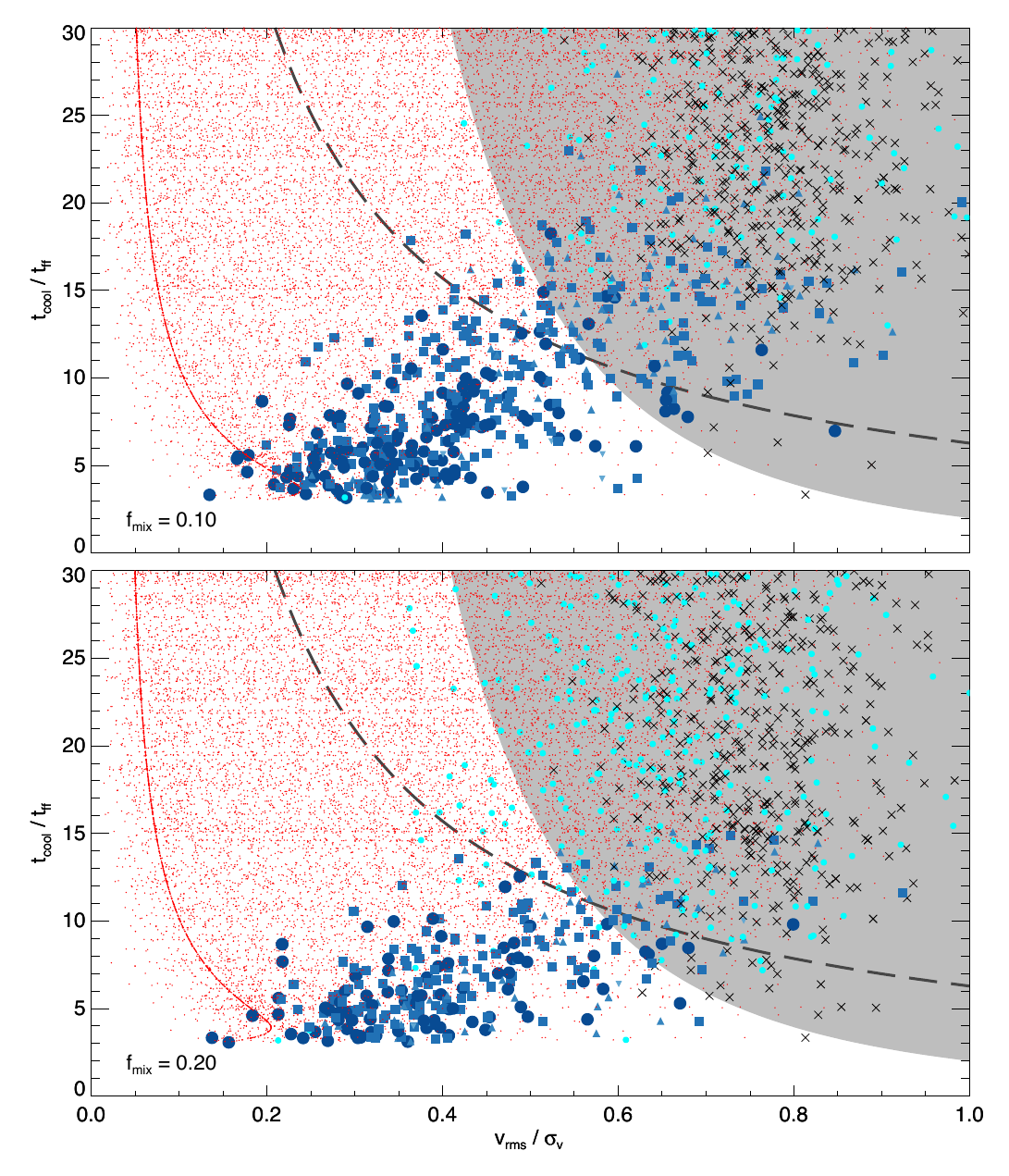} \\
\end{center}
\caption{ \footnotesize 
Dependence of condensation on both $t_{\rm cool}/t_{\rm ff}$ and velocity dispersion for $f_{\rm mix} = 0.1$ (top panel) and $f_{\rm mix} = 0.2$ (bottom panel) in trials of 1000 noise-driven gravity wave systems with randomly selected values of $t_{\rm cool}/t_{\rm ff}$.  In each set of trials for a given $t_{\rm cool}/t_{\rm ff}$ ratio, the gain factor begins with a small value and increases with each successive trial until a trial results in condensation.  Small red dots represent trials that fail to produce condensation by $t = 100 t_{\rm ff}(r_1)$.  Blue symbols represent trials that end in condensation, and the symbol type indicates the value of $L_{\rm path}$ at the moment of condensation:  large circles for $L_{\rm path}/r_1 < 10$, squares for $10 < L_{\rm path}/r_1 < 20$, triangles for $20 < L_{\rm path}/r_1 < 40$, inverted triangles for $40 < L_{\rm path}/r_1 < 80$, and small cyan circles for $L_{\rm path}/r_1 > 80$. Black crosses show trials that failed to produce condensation even with the gain factor set to its maximum value. Within the range $10 \lesssim t_{\rm cool} / t_{\rm ff} \lesssim 20$, the onset of condensation is at $v_{\rm rms} \approx (0.4$--$0.6) \sigma_v$ for $f_{\rm mix} = 0.1$. Setting $f_{\rm mix} = 0.2$ strongly suppresses condensation in systems with $t_{\rm cool} / t_{\rm ff} \gtrsim 15$, but the onset of condensation remains at $v_{\rm rms} \approx (0.4$--$0.6) \sigma_v$ for  $10 \lesssim t_{\rm cool} / t_{\rm ff} \lesssim 15$. 
The model assumption of volume-averaged thermal balance may be violated in the shaded region. 
Gray shading indicates the region of parameter space in which turbulent dissipation may violate the thermal-balance assumption by depositing heat faster than radiative cooling can shed it.  According to the model in \S \ref{sec-Coupling}, turbulent heating equals radiative cooling at the border of that region when the dissipation parameter is set to $\Gamma =1$, and it moves to the right for lower values of $\Gamma$.  The dashed charcoal line corresponds to the \citet{Gaspari_2018ApJ...854..167G} mixing criterion, as described in Appendix~\ref{sec-AppendixMixing}.
and the prominent ridge of red dots on the left corresponds to limit cycles without noise.   
\vspace*{1em}
\label{fig-sample}}
\end{figure*}

In general, the amount of noise required to induce condensation in a system with $t_{\rm cool} / t_{\rm ff} \approx 10$ corresponds to $0.4 \lesssim \sigma_{\rm t}/\sigma_v \lesssim 0.6$, and the critical velocity dispersion needs to be greater for larger values of $t_{\rm cool} / t_{\rm ff}$.  Figure~\ref{fig-sample} shows two numerical experiments designed to probe the relationship between a system's $t_{\rm cool} / t_{\rm ff}$ ratio and the amount of turbulence needed to drive condensation within it.  Each panel shows 1000 systems with randomly selected $t_{\rm cool} / t_{\rm ff}$ ratios in the range 3 to 30.  Symbols in the figure show the results of trials like those in Figures \ref{fig-blob_orbits_5} through \ref{fig-blob_orbits_20}.   The noise gain factor $\eta_0$ is small at the beginning of each set of trials for a given $t_{\rm cool} / t_{\rm ff}$ ratio, and $\eta_0$ increases with each successive trial until one ends in condensation. Small red dots show trials that failed to produce condensation before the end of the trial at $t = 100 t_{\rm ff}(r_1)$.   Blue symbols show trials that ended in condensation, with the symbol type representing the value of $L_{\rm path}$ at condensation.  Black crosses show trials that failed to produce condensation even with the noise gain factor set to its maximum value.  The prominent ridge of red points on the left side of each panel corresponds to low-noise oscillations on the limit cycle.

The top panel presents results for $f_{\rm mix} = 0.1$ and the bottom one presents results for $f_{\rm mix} = 0.2$. Both panels show that condensation in systems with $t_{\rm cool} / t_{\rm ff} \lesssim 10$ is relatively easy to stimulate with turbulence having $\sigma_{\rm t} \sim 0.3 \sigma_v$ on length scales $\sim r$.  However, condensation in systems with $10 \lesssim t_{\rm cool} / t_{\rm ff} \lesssim 20$ requires $\sigma_{\rm t} \sim 0.5 \sigma_v$, and the upper bound on $t_{\rm cool} / t_{\rm ff}$ among condensing systems is somewhat sensitive to mixing.   For example, raising the $f_{\rm mix}$ parameter from 0.1 to 0.2 eliminates systems with $t_{\rm cool} / t_{\rm ff} > 15$ that condense with $L_{\rm path}/r_1 < 80$.  The idealized oscillator model is unlikely to be valid for larger values of $L_{\rm path}/r_1$, because it implicitly assumes that an oscillating entropy perturbation remains coherent while passing through a column density of ambient gas many times greater than the column density of the perturbation itself.   Also, most of the points corresponding to condensation with $L_{\rm path}/r_1 > 80$ are within the shaded regions of Figure~\ref{fig-sample}, in which the assumption of thermal balance may be invalid because turbulent heating exceeds radiative cooling.  

These results are significant because galaxy cluster cores that harbor multiphase gas typically have $10 \lesssim \min (t_{\rm cool} / t_{\rm ff}) \lesssim 20$  \citep{Voit_2015Natur.519..203V,Hogan_2017_tctff} and $\sigma_{\rm t} / \sigma_v \approx 0.5$ \citep{McNamara_2014ApJ...785...44M,Russell_2016MNRAS.458.3134R,Hitomi_Perseus_2016Natur.535..117H}.  Without turbulence or large outflow velocities, buoyancy damping would suppress condensation in regions with $\alpha_K^{1/2} (t_{\rm cool} / t_{\rm ff}) \gtrsim 1$.  \citet{Voit_2017_BigPaper} therefore hypothesized that condensation beyond the isentropic center of such a system would require adiabatic uplift produced by bulk outflow, but the results of this section point to another possibility.  Turbulence stimulated by either AGN outflows or dynamical stirring by cosmological substructure may be what is causing condensation to happen in the region where $t_{\rm cool} / t_{\rm ff}$ reaches its minimum value, even in the presence of an entropy gradient with $\alpha_K \approx 2/3$.

\subsection{Next Steps}

Turbulent forcing of condensation should be commonplace in numerical simulations of mechanical AGN feedback, which are far less idealized than the heuristic model presented here.  The model's purpose is to assist with interpretations of those simulations and to clarify the properties that AGN feedback must have in order to self-regulate with properties similar to those of observed galaxy cluster cores.  This section's findings suggest that turbulence is essential for bringing about condensation in a medium with a non-negligible entropy gradient.  More sophisticated numerical simulations that solve complete sets of hydrodynamical and magnetohydrodynamical equations will be needed to quantify with greater precision the relationship between turbulence and the critical $t_{\rm cool} / t_{\rm ff}$ ratio for condensation within media with and without dynamically important magnetic fields.  Concurrently, simulations of self-regulating mechanical AGN feedback should be analyzed for correlations between the rate at which ambient gas condenses into cold clouds and the values of  $t_{\rm cool} / t_{\rm ff}$ and $\sigma_{\rm t} / \sigma_v$ in the ambient medium.  This paper will leave those investigations for the future and instead turn to the question of global self-regulation.

\section{Coupling between Precipitation, Feedback, and Turbulence}
\label{sec-Coupling}

Previous work on precipitation-regulated feedback has assumed that the condensation threshold depends on either the $t_{\rm cool} / t_{\rm ff}$ ratio alone \citep[e.g.,][]{McCourt+2012MNRAS.419.3319M,Sharma_2012MNRAS.420.3174S} or the parameter combination $\alpha_K^{1/2} (t_{\rm cool} / t_{\rm ff})$, which accounts for the failure of buoyancy damping to suppress condensation when $\alpha_K$ is sufficiently small \citep{Voit_2017_BigPaper}.  \footnote{ \citet{Gaspari_2017MNRAS.466..677G,Gaspari_2018ApJ...854..167G} have recently proposed a different criterion for condensation in a turbulent medium that compares $t_{\rm cool}$ with the timescale for eddy turnover.  See the Appendix for a discussion of how that criterion relates to the results of \S 3.} 
However, an additional dependence on $\sigma_{\rm t} / \sigma_v$, such as that shown in Figure~\ref{fig-sample}, can transform the feedback loop into a converging cycle that settles into a particular equilibrium state. This section explores global convergence to thermal balance in a precipitating medium by dropping the thermal balance requirement and developing a toy model that illustrates how turbulence can couple with precipitation.  The model illuminates the conditions under which coupling produces a long-lasting equilibrium state in agreement with observations of both the precipitation threshold and the turbulent velocity dispersion in galaxy cluster cores.

\subsection{A Toy Model for Coupling}

In order to keep the overall model as simple as possible, the dynamical timescales of the feedback system are assumed to be much shorter than its thermal and cosmological timescales, so that the ambient gas remains close to hydrostatic equilibrium.  The model further assumes that heat input does not change the slope of the ambient entropy profile, which is constrained to be $K(r) \propto r^{2/3}$ in a gravitational potential with $t_{\rm ff} \propto r$.  As a result, the gas temperature remains constant at $kT = 2 \mu m_p \sigma_v^2$, with a specific thermal energy $3 \sigma_v^2$ that remains invariant while the gas gains and loses heat energy.  Instead of causing the temperature to rise and fall, heat input causes the gas to expand and radiative losses cause the gas to contract, while gravitational work keeps $kT$ steady.  In such a constrained system, the $t_{\rm cool} / t_{\rm ff}$ ratio is the same at all radii at any given moment in time, with $t_{\rm cool} \propto n_e^{-1} \propto K^{3/2}$ at each radius.

The rate at which precipitation produces cold clouds at is taken to be $f_{\rm p} t_{\rm cool}^{-1}$ per unit gas mass, where $f_{\rm p}$ quantifies the condensation rate of cold clouds as a fraction of the cooling flow rate that would transpire if heating were absent.  According to the analysis of \S \ref{sec-Heuristic}, the condensation fraction $f_{\rm p}$ depends on both $t_{\rm cool} / t_{\rm ff}$ and $\sigma_{\rm t} / \sigma_v$.  Figure~\ref{fig-sample} shows that $f_{\rm p}$ rises sharply from $f_{\rm p} \ll 1$ to $f_{\rm p} \sim 1$ as $t_{\rm cool}$ decreases through a narrow range in $t_{\rm cool} / t_{\rm ff}$ and that the critical range moves to larger values of $t_{\rm cool} / t_{\rm ff}$ as $\sigma_{\rm t} / \sigma_v$ increases.  Precipitation-regulated systems with this feature therefore come into equilibrium at a $t_{\rm cool} / t_{\rm ff}$ value that depends on $\sigma_{\rm t} / \sigma_v$.  

Feedback fueled by condensation is assumed to restore specific energy to the ambient gas at a rate $\epsilon_H c^2 f_{\rm p} t_{\rm cool}^{-1}$, where $\epsilon_H$ is a feedback efficiency parameter that accounts for both the efficiency of accretion onto the central black hole and the fraction of accreted mass-energy that returns to the gas.  The model assumes that a fraction $f_{\rm t}$ of the returning energy goes into turbulence, and the remaining $1 - f_{\rm t}$ goes directly into heat.  Turbulent energy then decays into heat at a rate $\Gamma \sigma_{\rm t} / \sigma_v t_{\rm ff}$, where $\Gamma$ is a dimensionless free parameter that permits tuning of the dissipation rate and should be of order unity when the outer scale for driving of turbulence is $\sim r$.

Evolution of the turbulent velocity dispersion is therefore governed by
\begin{equation}
 \frac {d} {dt} \left( \frac {3} {2} \sigma_{\rm t}^2 \right) =
 	\frac {f_{\rm t} f_{\rm p} \epsilon_H c^2}  {t_{\rm cool}} 
	- \left( \frac {3} {2} \sigma_{\rm t}^2 \right) \frac {\Gamma \sigma_{\rm t}} {t_{\rm ff} \sigma_v}
	\label{eq-dsigmat_dt}
	\; \; ,
\end{equation}
assuming isotropic turbulence.  A similar expression for evolution of the ambient cooling time,
\begin{equation}
  \frac {dt_{\rm cool}} {dt} = 
  		(1 - f_{\rm t})  \frac {f_{\rm p} \epsilon_H c^2} {2 \sigma_v^2} 
			+ \frac {3\Gamma} {4} \frac {t_{\rm cool}} {t_{\rm ff}} 
			  \left( \frac {\sigma_{\rm t}} {\sigma_v} \right)^3 
			- \frac {3} {2} 
			\label{eq-dtcool_dt}
	\; \; ,
\end{equation}
follows from the relation $d \ln t_{\rm cool} / dt = d \ln K^{3/2} /dt$ for gas at a constant temperature (see the Appendix).  Notice that the equilibrium state of the system corresponds to 
\begin{equation}
 \frac {\sigma_{\rm t}} {\sigma_v} = \left( \frac {2 f_{\rm t}} {\Gamma} \right)^{1/3}
 							\left( \frac {t_{\rm cool}} {t_{\rm ff}} \right)^{-1/3}
	\label{eq-sigma_eq}
\end{equation}
and
\begin{equation}
  f_{\rm p} \left( \frac {t_{\rm cool}} {t_{\rm ff}} ,
  			\frac {\sigma_{\rm t}} {\sigma_v} \right) 
	        = f_0
  	        \equiv \frac {3 \sigma_v^2} {\epsilon_H c^2} 
	   \label{eq-f0}
     \; \; .
\end{equation}
Equation (\ref{eq-sigma_eq}) defines a locus in the $\sigma_{\rm t} / \sigma_v$ versus $t_{\rm cool} / t_{\rm ff}$ plane of Figure~\ref{fig-sample} along which dissipation of turbulence matches the rate at which feedback generates it in a thermally balanced system.   Equation (\ref{eq-f0}) defines $f_0$ to be the value of $f_{\rm p}$ at which feedback energy input fueled by precipitation balances radiative cooling.  Gray shading in the figure shows how the position of the locus defined by Equation (\ref{eq-sigma_eq}) depends on the factor $2 f_{\rm t} / \Gamma$.  Toward the right of the locus, dissipation of turbulence exceeds the generation rate when $f_{\rm p} = f_0$.  Toward the left, generation of turbulence exceeds dissipation when $f_{\rm p} = f_0$.

\begin{figure*}[t]
\begin{center}
\includegraphics[width=6.5in, trim = 0.0in 0.2in 0.0in 0.0in]{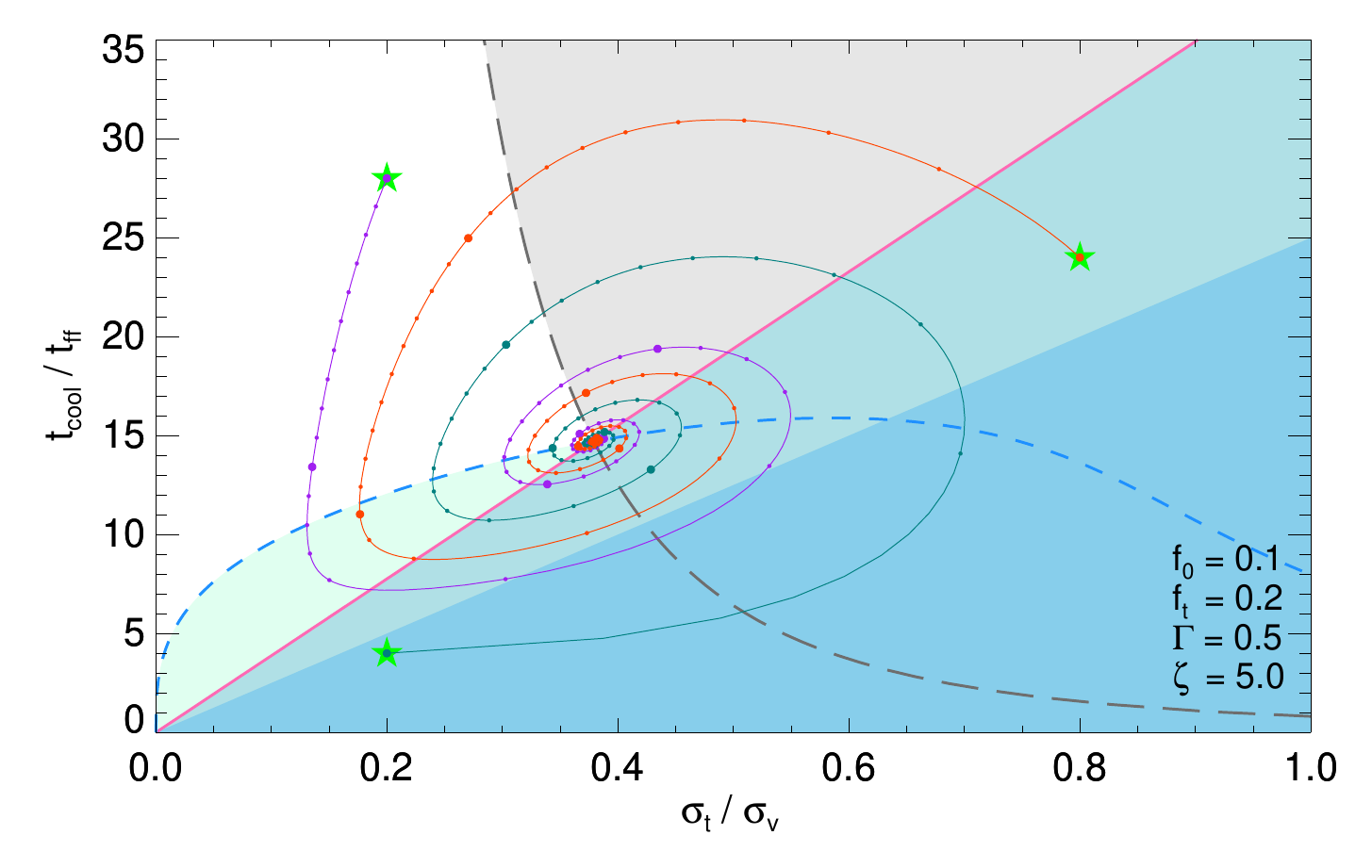} \\
\end{center}
\caption{ \footnotesize 
Convergence of an idealized precipitation-regulated system to its equilibrium state.  That equilibrium state is at the intersection of the long-dashed gray line defined by Equation (\ref{eq-sigma_eq}) and the bold pink line, along which the total feedback heating rate fueled by precipitation equals the total radiative cooling rate (i.e. $f_{\rm p} = f_0$).  Each of the three green stars marks the starting point of a trajectory resulting from numerical integration of Equations (\ref{eq-dsigmat_dt}) and (\ref{eq-dtcool_dt}) given the parameter values $f_0 = 0.1$, $f_{\rm t} = 0.2$, $\Gamma = 0.5$, $\zeta = 5$, with $f_{\rm p}$ defined by Equation (\ref{eq-fp}).  Small filled circles on each trajectory indicate time steps equal to $t_{\rm ff}$, and large filled circles indicate steps of $10 t_{\rm ff} \sim t_{\rm cool}$.  A short-dashed blue line shows the locus along which stimulation of turbulence equals dissipation of turbulence when the system is out of equilibrium.  In the white region cooling exceeds heating while turbulence decays, and so the trajectory proceeds counterclockwise around the equilibrium point.  When the trajectory crosses the short-dashed blue line into the lightest blue region, precipitation produces enough feedback to boost turbulence.  In the regions with deeper blue shading, feedback energy input exceeds cooling, and $f_{\rm p} \geq 0.5$ in the darkest blue region.  Dissipation of turbulence starts to exceed stimulation upon the next crossing of the short-dashed blue line, and turbulence subsequently declines while the system continues to heat.  At the next crossing of the pink line, into the gray region, cooling starts to exceed total energy input, but decay of residual turbulence can still cause net heating.  For this parameter set, $f_{\rm t}$ satisfies the condition in Equation (\ref{eq-convergence}), and so all trajectories converge to a fixed point at $t_{\rm ff}/t_{\rm cool} = 14.7$, $\sigma_{\rm t}/\sigma_v = 0.38$.
\vspace*{1em}
\label{fig-cycle}}
\end{figure*}

Figure \ref{fig-cycle} shows that the locus of turbulent balance crosses the precipitation threshold in the vicinity of $\sigma_{\rm t} / \sigma_v \approx 0.4$ and $10 \lesssim t_{\rm cool} / t_{\rm ff} \lesssim 20$, in agreement with observations of galaxy cluster cores, when $f_{\rm t}$ and $\Gamma$ are both of order unity.  Deriving a more precise equilibrium point for the system requires better knowledge of the function $f_{\rm p} ( t_{\rm cool} / t_{\rm ff} , \sigma_{\rm t} / \sigma_v )$ than Figure~\ref{fig-sample} provides.  Trying to push the toy model any further than this might not be valid.  However, it is still useful for illustrating, at least qualitatively, how such a feedback system converges toward equilibrium.

\subsection{Convergence toward Self-regulating Equilibrium}

Non-equilibrium evolution of the system's configuration in the plane of Figure~\ref{fig-cycle} can be computed from Equations (\ref{eq-dsigmat_dt}) and (\ref{eq-dtcool_dt}), with the assistance of an approximation for the function $f_{\rm p} ( t_{\rm cool} / t_{\rm ff} , \sigma_{\rm t} / \sigma_v )$.  The following approximation corresponds qualitatively to the surface density of blue circles in the top panel of Figure~\ref{fig-sample}:
\begin{equation}
  f_{\rm p} = \frac {P^\zeta} {P^{\zeta} + 1} 
  \; \; \; , \; \; \; 
  P  =  25 \frac { {\sigma_{\rm t}} / {\sigma_v}  } 
                        { t_{\rm cool} / t_{\rm ff} } 
       \; \; \; . 
       \label{eq-fp}
\end{equation}
The prefactor of 25 was chosen so that $P = 1$ and $f_{\rm p} = 0.5$ along a line containing the configuration points ($\sigma_{\rm t} / \sigma_v$,$t_{\rm cool} / t_{\rm ff}$) = (0.2,5), (0.4,10), and (0.8,20), which runs along the transition from noncondensing to condensing trajectories in the top panel of Figure \ref{fig-sample}.  The parameter $\zeta$ determines the sharpness of the transition to condensation as $\sigma_{\rm t} / \sigma_v$ rises at constant $t_{\rm cool} / t_{\rm ff}$ and has a fiducial value $\zeta = 5$ in this paper's computations.  

\begin{figure*}[t]
\begin{center}
\includegraphics[width=6.5in, trim = 0.0in 0.2in 0.0in 0.0in]{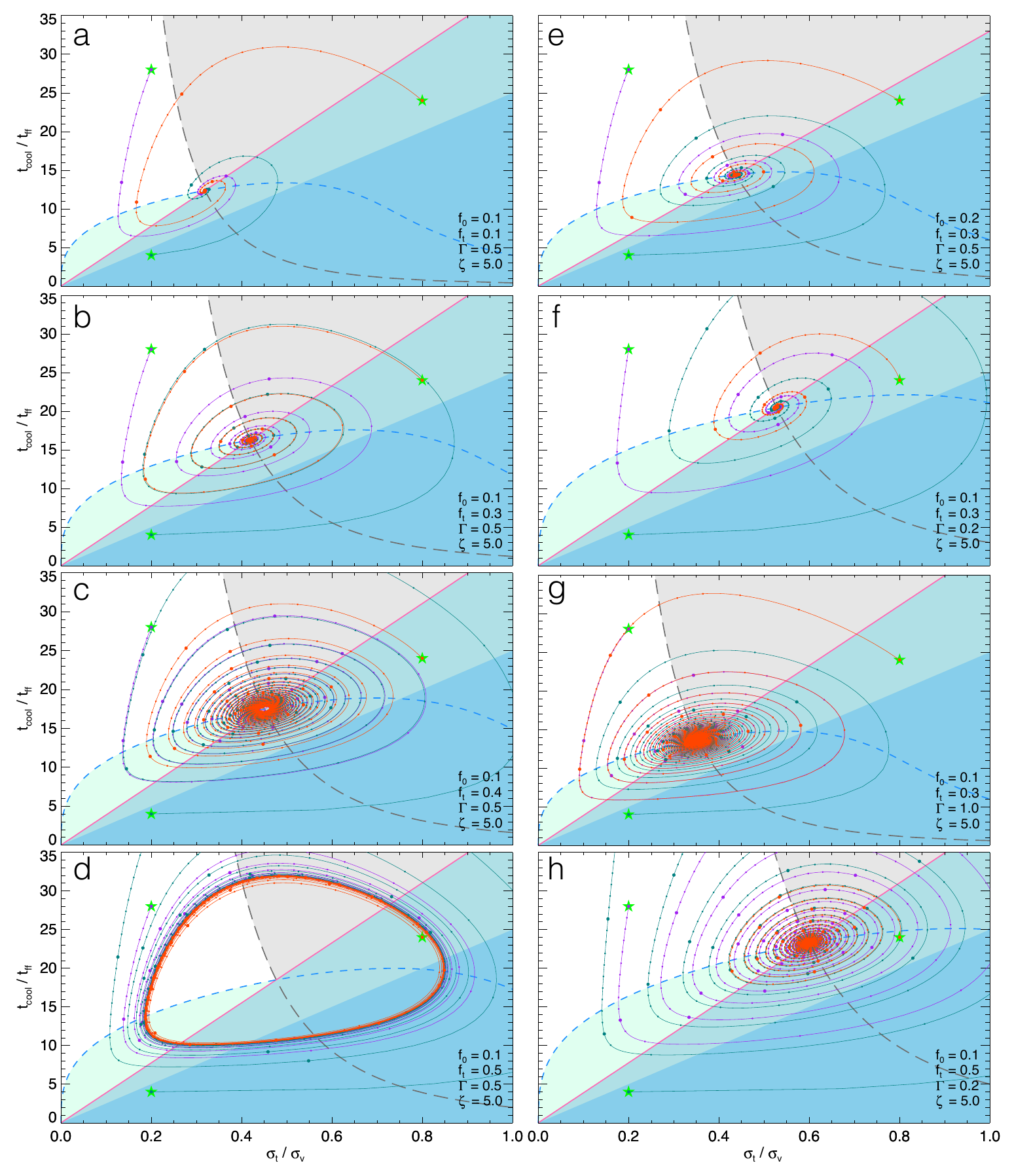} \\
\end{center}
\caption{ \footnotesize 
Convergence to equilibrium in differing precipitation-regulated systems.  Lines, points, and shading in each panel are the same as in Figure \ref{fig-cycle}.   Labels in each panel show the parameter settings for that panel, and identical initial conditions are used for the three trajectories in each panel.  In panels (a) through (d), the only system parameter that changes is $f_{\rm t}$, the fraction of feedback energy input that goes into turbulence, with $f_{\rm t}$ = 0.1, 0.3, 0.4, and 0.5, respectively.  In this sequence of panels, the fixed-point solution to Equations (\ref{eq-dsigmat_dt}) and (\ref{eq-dtcool_dt}) becomes unstable as $f_{\rm t}$ increases through 0.35.  The asymptotic trajectory becomes a limit cycle for larger values of $f_{\rm t}$, with greater turbulent energy input corresponding to greater excursions in both $t_{\rm cool} / t_{\rm ff}$ and $\sigma_{\rm t} / \sigma_v$.  Comparing panels (b) and (e) shows that raising the value of $f_0$ required for equilibrium moves the equilibrium point to lower $t_{\rm cool}/t_{\rm ff}$ and greater $\sigma_{\rm t}/\sigma_v$.  
Comparing panels (b), (f), and (g) shows that lowering $\Gamma$ (equivalent to increasing the injection scale for turbulence) moves the equilibrium point farther from the origin and speeds convergence to equilibrium.  Comparing panels (d) and (h) shows that lowering $\Gamma$ can enable the system to converge to an equilibrium fixed point even for $f_{\rm t} = 0.5$.
\vspace*{1em}
\label{fig-cycles}}
\end{figure*}

Trajectories in Figure \ref{fig-cycle} show the results of integrating Equations (\ref{eq-dsigmat_dt}) and (\ref{eq-dtcool_dt}) for the parameter values $f_0 = 0.1$, $f_{\rm t} = 0.2$, $\Gamma = 0.5$, and $\zeta = 5$.  The fiducial parameter values are inspired by the primary simulation of \citet{Li_2015ApJ...811...73L}, in which the time-averaged precipitation rate is $\approx 10$\% of the pure cooling rate and the time-averaged turbulent dissipation rate is $\approx 20$\% of the total heat input \citep{Li_2017ApJ...847..106L}. Starting at low $\sigma_{\rm t}$ and long $t_{\rm cool}$ gives the purple trajectory,  starting at low $\sigma_{\rm t}$ and short $t_{\rm cool}$ gives the teal trajectory, and starting at high $\sigma_{\rm t}$ and long $t_{\rm cool}$ gives the red trajectory.  All of the trajectories become similar after a couple of cooling times and eventually converge to an equilibrium point with $t_{\rm ff}/t_{\rm cool} = 14.7$ and $\sigma_{\rm t}/\sigma_v = 0.38$.
 
\begin{table}
\centering
\caption{Coupling Model Parameters}
\begin{tabular}{cl}
\hline
 ~ & ~ \vspace{-0.5em} \\
$f_0$ & Equilibrium condensation fraction \\
$f_{\rm t}$  & Fraction of feedback energy going into turbulence \\
$\Gamma$  & Turbulent dissipation parameter ($\sim r/L$) \\
$\zeta$  & Precipitation threshold sharpness parameter \vspace*{0.5em}  \\ 
\hline
\end{tabular}
\end{table}

Convergence to equilibrium depends critically on $f_{\rm t}$, the fraction of feedback energy going into turbulence, as shown by panels (a) through (d) of Figure \ref{fig-cycles}.  If all other system parameters from Figure \ref{fig-cycle} are held constant, the asymptotic trajectories for $f_{\rm t} \gtrsim 0.3$ become limit cycles, with amplitudes in $t_{\rm cool} / t_{\rm ff}$ and $\sigma_{\rm t} / \sigma_v$ that depend sensitively on $f_{\rm t}$.  The condition that determines whether the system converges to a fixed point or a limit cycle can be obtained by linearizing Equations (\ref{eq-dsigmat_dt}) and (\ref{eq-dtcool_dt}), combining them into a single, second-order oscillator equation, and evaluating the sign of the damping term.  According to that calculation, the system converges to a fixed point for
\begin{equation}
  f_{\rm t} <  -  \frac {3x_0^2 D_y}  {2(D_x - 3) - 3x_0^2 (D_y - 1)}
  \label{eq-convergence}
\end{equation}
where $x_0$ is the value of $\sigma_{\rm t} / \sigma_v$ at the equilibrium point and the sensitivity of $f_{\rm p}$ is parameterized by $D_x \equiv  \partial \ln f_{\rm p} / \partial \ln \sigma_{\rm t}$ and  $D_y \equiv  \partial \ln f_{\rm p} / \partial \ln t_{\rm cool}$.  The form of $f_{\rm p}$ in Equation (\ref{eq-fp}) gives $D_x \approx - D_y \approx \zeta$, which leads to the convergence criterion $f_{\rm t} \lesssim 0.35$ for $\zeta = 5$ and $x_0 = 0.4$, in accordance with the left panels of Figure \ref{fig-cycles}.

More generally, if the transition to condensation happens within a narrower region of the $\sigma_{\rm t}$--$t_{\rm cool}$ plane, then the absolute values of $D_x$ and $D_y$ substantially exceed unity.  And if $f_{\rm p}$ depends only on the ratio of $\sigma_{\rm t} / \sigma_v$ to $t_{\rm cool} / t_{\rm ff}$, as in Equation (\ref{eq-fp}), then $D_x = - D_y$, in which case the convergence criterion approaches $f_{\rm t} < 3x_0^2 / (2 + 3x_0^2)$ as the transition region becomes increasingly narrow.  Using Equation (\ref{eq-sigma_eq}) to represent $x_0$ in terms of $t_{\rm cool} / t_{\rm ff}$, $\Gamma$, and $f_{\rm t}$ then yields the approximate convergence criterion
\begin{equation}
  f_{\rm t} \lesssim \frac {3} {2} \left( \frac {2} {\Gamma} \right)^2 
  			     \left( \frac {t_{\rm cool}} {t_{\rm ff}} \right)_0^{-2}
	\label{eq-approx_convergence}
\end{equation}
in the limit $x_0^2 \ll 1$, with the zero subscript denoting the equilibrium value of $t_{\rm cool} / t_{\rm ff}$.  

The approximation in Equation (\ref{eq-approx_convergence}) is not as precise as Equation (\ref{eq-convergence}) but reveals the inverse relationship between $\Gamma$ and the convergence threshold for $f_{\rm t}$.  Comparing panels (b), (f), and (g) in Figure \ref{fig-cycles} shows that decreasing $\Gamma$ does indeed speed convergence to a fixed point.  Reducing $\Gamma$ corresponds to a smaller damping rate and implies a larger scale length for stimulation of turbulence with a velocity dispersion $\sigma_{\rm t}$.  Lowering the value of $\Gamma$ therefore lengthens the timescale for decay of turbulence and diminishes the decrease in $\sigma_{\rm t}$ that can occur during the cooling-dominated portion of the feedback cycle.  As a result, convergence requires fewer cycles for smaller $\Gamma$.  Figure \ref{fig-cycles} panel pair (d)-(h) illustrates how lowering $\Gamma$ allows the convergence condition to be satisfied for a larger value of $f_{\rm t}$.

\subsection{Interpretation of the Toy Model}

The toy model defined by Equations (\ref{eq-dsigmat_dt}) and (\ref{eq-dtcool_dt}) is clearly a vast oversimplification of the precipitation-regulated feedback loops realized in numerical simulations, not to mention real galaxies.  It is also internally inconsistent, because it exhibits cyclical behavior with a period that is constrained to be independent of radius, even though the natural timescale for the cycle (i.e. $t_{\rm cool}$) increases with radius.  Nevertheless, the model exhibits several qualitative features that are probably generic to precipitation-regulated systems of greater complexity:
\begin{enumerate}

\item  If precipitation-regulated feedback generates turbulence, then it can enable condensation in ambient media with $t_{\rm cool} / t_{\rm ff}$ values larger than the precipitation threshold in a static system.

\item  The nonequilibrium evolution of a turbulent precipitation-regulated feedback system proceeds quasi-cyclically with a period similar to the cooling time.  Typically, the ambient cooling time in such a system rises with increasing radius, implying that cycles of nonequilibrium evolution proceed more slowly at greater radii.  The dominant time scale for cycling is likely to be the cooling time at the innermost radius at which $t_{\rm cool} / t_{\rm ff}$ is minimized, because that is where the ambient medium is most susceptible to condensation.  

\item  A precipitation-regulated system can converge to a well-regulated equilibrium state if a relatively small proportion of the feedback energy goes into turbulence.  That equilibrium state has the following properties:  (1) turbulence assists precipitation by suppressing buoyancy damping, (2) feedback energy input fueled by precipitation is able to match radiative cooling, and (3) generation of turbulence by feedback balances dissipation of turbulence.  

\item  For physically reasonable ranges in its parameter space, the toy model self-regulates with $10 \lesssim t_{\rm cool} / t_{\rm ff} \lesssim 20$ and $0.3 \lesssim \sigma_{\rm t} / \sigma_v \lesssim 0.6$.  More generally, the range of self-regulating equilibria depends on the form of the function $f_{\rm p} ( t_{\rm cool} / t_{\rm ff} , \sigma_{\rm t} / \sigma_v )$, which describes how the condensation rate depends on turbulence.

\item  Precipitation-regulated feedback that puts a large proportion of the feedback energy into turbulence is unable to converge to a well-regulated equilibrium state.  Instead, it produces an indefinite number of sporadic feedback bursts.  Panel (d) of Figure \ref{fig-cycles}, in which half the feedback energy goes into turbulence, shows an example.  With the parameter set $f_0 = 0.1$, $f_{\rm t} = 0.5$, $\Gamma = 0.5$, $\zeta = 5$, the asymptotic state of the idealized system is a limit cycle that briefly passes through a phase with a precipitation rate comparable to the pure cooling rate  and subsequently spends a much larger fraction of the cycle with $t_{\rm cool} / t_{\rm ff} > 20$ and a negligible precipitation rate.  
A time period similar to the maximum value of $t_{\rm cool}$ reached on the limit cycle must pass before cooling of the ambient medium can lead to more precipitation.  In a less idealized system, this type of behavior would manifest as a series of sporadic star-formation outbursts, along with temporally correlated AGN outbursts of shorter duration.

\end{enumerate}

These features of the toy model collectively provide a valuable context for classifying and analyzing the behavior of the feedback mechanisms that regulate star formation in galaxies. Numerical simulations will be needed to determine how well the toy model captures the generic features of convergence to self-regulated equilibrium, as well as the circumstances under which feedback is episodic rather than continuous.  Among the observationally testable results will be predictions for correlations among three key observables:  (1) the minimum value of $t_{\rm cool} / t_{\rm ff}$ in the hot ambient medium, (2) the turbulent velocity dispersion in each multiphase gas component as a fraction of the local stellar velocity dispersion (i.e. $\sigma_{\rm t} / \sigma_v$), and (3) the mass distribution of stellar populations with ages $\lesssim t_{\rm cool}$.  Stellar population ages will be crucial for determining the amplitude of a feedback cycle and the time interval that has passed since the last maximum in the precipitation rate, because that information will determine the phase of the feedback cycle in a configuration space similar to Figure \ref{fig-cycle} and the resulting correlations of stellar population properties with $\min(t_{\rm cool} /t_{\rm ff})$ and $\sigma_{\rm t} / \sigma_v$.

\section{Concluding Thoughts}
\label{sec-Concluding}

This paper has developed a simple heuristic model to assess the impact of turbulence on precipitation in a medium with $t_{\rm cool} / t_{\rm ff} \gg 1$ and $K \appropto r^{2/3}$.  Its main finding is that turbulent forcing can stimulate precipitation in a medium with $10 \lesssim t_{\rm cool} / t_{\rm ff} \lesssim 20$ by raising the velocity dispersion of the ambient medium to $\approx 0.5 \sigma_v$, which is similar to the velocity dispersions observed in multiphase galaxy cluster cores.   It also finds that the level of turbulence needed to stimulate precipitation is positively correlated with $t_{\rm cool} / t_{\rm ff}$ and that it is possible for coupling between turbulence and precipitation in such a feedback system to promote convergence to a well-regulated equilibrium state.

It arrives at those findings by representing entropy perturbations as internal gravity waves in an entropy-stratified medium.  Those waves are thermally unstable in a thermally balanced background medium, and their amplitudes steadily grow in the linear regime.  Growth saturates at an Eulerian entropy amplitude $\delta \ln K \approx \alpha_K^{1/2} (t_{\rm ff} / t_{\rm cool})$ when the rate of energy transfer into dissipative acoustic modes equals the rate at which thermal pumping adds energy to gravity waves.  Dissipation therefore prevents condensation in an otherwise static medium with $t_{\rm cool} / t_{\rm ff} \gg 1$.  The trajectories of entropy perturbations consequently converge onto a limit cycle in the $r$--$K$ plane (see Figure \ref{fig-Linear_Damped}).

Nonlinearity of the gravitational potential causes the limit cycle's centroid to drift to lower entropy and smaller radius if it is uncompensated by additional heating (see Figure \ref{fig-Nonlinear_Nodrift}).  Drift occurs because a thermally cycling perturbation spends slightly more time at larger radii, in a net cooling state, than at smaller radii, in a net heating state.  Drift can be counterbalanced with additional heat, but the heating correction depends on the limit cycle's amplitude, and consequently on the perturbation's wavelength.  It is therefore not possible for a single heating correction to counterbalance entropy drift across the entire spectrum of gravity wave modes.  The calculations in this paper opt to counterbalance long-wavelength ($kr = 1$) modes, because those are the most consequential for precipitation-driven feedback.

Turbulent mixing is suppressed in most of this paper's calculations.  When mixing is included, it is parameterized in terms of the ratio $f_{\rm mix} = \omega_{\rm mix} / \omega_{\rm D}$, which specifies the mixing rate as a fraction of the rate at which the perturbation's kinetic energy dissipates.  Naively, one expects $f_{\rm mix}$ to be of order unity, with $\omega_{\rm mix} \sim k \sigma_{\rm t}$, which is similar to the inverse of the eddy turnover time.  That expectation has led to proposals that the critical criterion for condensation is for $t_{\rm cool}$ to be less than the eddy turnover time \citep[e.g.,][]{BanerjeeSharma_2014MNRAS.443..687B,Gaspari_2017MNRAS.466..677G,Gaspari_2018ApJ...854..167G}.  However, such a criterion implies that slower turbulence would allow condensation to happen for a larger ambient value of $t_{\rm cool}/t_{\rm ff}$.  (See the Appendix for a more detailed discussion of the eddy turnover time.) 

This paper's findings point in a complementary direction, implying that an increase in turbulence is needed in order to compensate for the increase in buoyancy damping that goes along with an increase in $t_{\rm cool}/t_{\rm ff}$.  However, too large an increase in turbulence can halt condensation if dissipation of turbulence into heat exceeds the radiative cooling rate. Controlled numerical simulations of turbulent condensation of an entropy-stratified medium in a gravitational potential will therefore be needed to determine the conditions under which an increase in turbulence promotes condensation by counteracting buoyancy or suppresses it through turbulent mixing.  The results may well depend on how effectively magnetic fields suppress mixing \citep[see also][]{JiOhMcCourt_2017arXiv171000822J} and on whether a perturbation's wavelength exceeds the outer scale for driving of turbulence.

If mixing is inconsequential, then perturbations with sufficiently large outward velocities can progress to condensation if they reach maximum altitude with an Eulerian entropy contrast large enough to force the gravity wave oscillator into overdamping (see Figure \ref{fig-Ballistic}).  That happens when the cooling time {\em local to the perturbation} is approximately equal to the freefall time from its current location.  Oscillations then cease because the perturbation condenses faster than it can descend to a layer of equivalent entropy.  However, the outward kicks needed to produce condensation in a static ambient medium with $t_{\rm cool} / t_{\rm ff} \gtrsim 10$ can exceed the medium's sound speed.

Turbulence in a system with these characteristics provides an easier route to condensation.  In the context of a simple gravity wave oscillator, turbulent momentum impulses are equivalent to forcing with Markovian noise. When noise is introduced, a perturbation's entropy amplitude diffuses around its limit cycle.  If that random walk in amplitude reaches the threshold for overdamping, then the perturbation is destined to condense (see Figures \ref{fig-blob_orbits_5}, \ref{fig-blob_orbits_10}, and \ref{fig-blob_orbits_20}).  In an ambient medium with $10 \lesssim t_{\rm cool} / t_{\rm ff} \lesssim 20$, the critical amount of diffusion is equivalent to $\sigma_{\rm t} \approx 0.5 \sigma_v$, with greater $t_{\rm cool} / t_{\rm ff}$ requiring greater $\sigma_{\rm t} / \sigma_v$ (see Figure \ref{fig-sample}).

If the threshold value of $t_{\rm cool} / t_{\rm ff}$ for condensation does indeed correlate with $\sigma_{\rm t} / \sigma_v$, then turbulence couples interestingly with precipitation.  Convergence to a well-regulated equilibrium state is possible if the fraction of feedback energy going into turbulence is not too large (see Figure \ref{fig-cycle}).  However, precipitation-regulated systems with a large fraction of feedback energy going into turbulence cannot settle into a steady equilibrium state (see Figure \ref{fig-cycles}).  Instead, they undergo periodic outbursts of condensation and feedback, interspersed with much longer intervals of relative quiescence.

The relationships between these simple models and real galactic feedback systems can be tested by looking for correlations among $\min (t_{\rm cool} / t_{\rm ff})$, $\sigma_{\rm t} / \sigma_v$, and the prominence of young stellar populations.  Numerical simulations of precipitation-regulated feedback processes will be needed to specify particular observational tests, because the analytical models presented here are too simplistic for that purpose.  But their simplicity is a virtue for the purpose of providing an overall context for studying and analyzing the relationships between precipitation, turbulence, and feedback in the circumgalactic media of massive galaxies.

\vspace*{0.5em}

I would like to thank Steve Balbus, Greg Bryan, Megan Donahue, Gus Evrard, Max Gaspari, Yuan Li, Mike McCourt, Chris McKee, Norm Murray, Brian McNamara, Paul Nulsen, Brian O'Shea, Peng Oh, and Prateek Sharma for helpful conversations.  This research was supported in part by the National Science Foundation under grant no.~NSF PHY-1125915, because its author benefitted from the Kavli Institute for Theoretical Physics through their workshop on the Galaxy-Halo Connection.  

\appendix


\section{Simplified Equations of Motion for a Nonadiabatic Internal Gravity Wave}

Section 3 of the paper presents a dynamical analysis of an oscillator governed by Equations (5) and (6).  Those equations represent a reduction of the fluid equations to a system with only two degrees of freedom that mimics a nonadiabatic internal gravity wave.  Equation (5) follows from the momentum equation:
\begin{equation}
  \frac {d {\bf v}} {dt} = - \frac {\nabla P} {\rho}  - \nabla \phi
    \; \; .
\end{equation}
For a perturbation to a background state in hydrostatic equilibrium, the momentum equation becomes
\begin{equation}
    \left(1 + \delta_\rho \right) \frac {d {\bf v}} {dt} = \left( \delta_P- \delta_\rho \right) \nabla \phi 
  					- \frac {\bar{P}} {\bar{\rho}} \nabla \delta_P
    \; \; ,
\end{equation}
where $\delta_P \equiv (P - \bar{P}) / \bar{P}$ and $\delta_\rho \equiv (\rho - \bar{\rho}) / \bar{\rho}$ and quantities with bars represent the background state.  The only nonlinear term is $\delta_\rho (d {\bf v} / {dt})$, representing the change in inertia associated with the perturbation.

In gas with an equation of state $P \propto K \rho^{5/3}$, the linearized momentum equation can be written as
\begin{equation}
  \frac {d {\bf v}} {dt} = \left( \frac {3} {5} \delta_K + \frac {2} {5} \delta_P \right)  \nabla \phi 
  					- \frac {3} {5} c_s^2 \nabla \delta_P
    \; \; ,
\end{equation}
where $\delta_K \equiv (K - \bar{K}) / \bar{K}$, and $c_s$ is the sound speed in the unperturbed gas.  The radial part of this equation reduces to 
\begin{equation}
  \dot{\Delta}_v = \left( \frac {3} {5} \delta_K + \frac {2} {5} \delta_P \right)  \frac {2} {t_{\rm ff}} 
  					- \frac {3} {5} \frac {c_s^2} {\sigma_v^2}  
					  \left( \frac {\partial \ln \delta_P} {\partial \ln r} \right)
					  \frac {\delta_P} {t_{\rm ff}}
					\; \; ,
\end{equation}
where $\Delta_v \equiv v_r / \sigma_v$ in an isothermal potential (constant $\sigma_v$) with gravitational acceleration $2 \sigma_v^2 / r = 2 \sigma_v / t_{\rm ff}$.   Internal gravity waves with frequencies much less than those of sound waves of equivalent wavelength have $\delta_K \gg \delta_P$, meaning that the $2 \delta_P / 5$ term can be neglected.  

The remaining term containing $\delta_P$ may still be significant if $\partial \ln \delta_P / \partial \ln r$ is large, and it affects the frequency of an internal gravity wave by coupling radial and tangential motions via the tangential momentum equation and the continuity equation.  In a linear analysis, this coupling produces the $k_\perp^2 / k^2$ factor in Equation (1), which accounts for the tangential inertia of a gravity wave with a nonzero wavenumber in the radial direction \citep[see, e.g.,][]{bnf09}.  Dropping this term gives 
\begin{equation}
  \dot{\Delta}_v = \frac {6} {5 t_{\rm ff}} \delta_K
  	\; \; ,
\end{equation}
which represents the motion of purely tangential internal gravity waves in the linear regime.  

In an entropy-stratified adiabatic medium with $\alpha_K \equiv d \ln \bar{K} / d \ln r$, the entropy equation is $\dot{\delta}_K = - \alpha_K \Delta_v / t_{\rm ff}$, resulting in simple harmonic oscillations of frequency $\omega_{\rm buoy} = (6 \alpha_K / 5)^{1/2} t_{\rm ff}^{-1}$.  However, a non-adiabatic medium can be thermally unstable if the entropy contrast of a stationary perturbation grows as $\omega_{\rm ti} \delta_K$, with $\omega_{\rm ti} > 0$.  In that case, the appropriate entropy equation is Equation (6) with $\omega_{\rm mix} = 0$, and thermal instability pumps exponential growth of internal gravity waves on a time scale $\sim t_{\rm cool}$.

Section 3 explores what happens to these pumped harmonic oscillations when their amplitudes become large.  Generic dissipation of kinetic wave energy into the ambient medium at a rate $\omega_{\rm D}$ can be modeled by adding the term $- \omega_{\rm D} \Delta_v$ to the right-hand side of the equation for $\dot{\Delta}_v$.  Drag forces on an isolated buoyant blob of size $l$ cause its kinetic energy to dissipate with $\omega_D \sim | \Delta_v | \sigma_v / l  \sim | \Delta_v | (r/l) \omega_{\rm buoy}$.  However, the concept of drag is less satisfactory for understanding how the kinetic energy of a wave-like disturbance dissipates.  In lieu of a conceptually questionable invocation of drag, \citet{Voit_2017_BigPaper} show that nonlinear wave coupling through the $\delta_\rho (d {\bf v} / dt)$ term in the momentum equation (i.e. buoyancy damping) transfers gravity wave energy into other wave modes at a rate $\omega_D \sim | \Delta_v | (kr) \omega_{\rm buoy}$, which closely resembles the dissipation rate for a blob experiencing drag.  Equation (5) is therefore suitable for modeling, at least qualitatively, damped oscillations of either a bobbing blob of a particular size or a tangential internal gravity wave with a specific wavelength, given an appropriate expression for $\omega_{\rm D}$.

\section{Discrete Langevin Noise}

The $\eta$ term in Equation (5) drives the gravity wave oscillator via a series of random Markovian impulses $\{ \eta_i \}$ with $\langle \eta \rangle = 0$.  In the absence of gravity, the value $\Delta_{i+1}$ of $\Delta_v$ at time $t_{i+1}$ depends on its value $\Delta_i$ at time $t_i = t_{i+1} - \Delta t$ according to
\begin{equation}
  \Delta_{i+1} = (1 - \omega_{\rm D} \Delta t ) \Delta_i + \eta_i \Delta t
  \; \; .
\end{equation}
The variance of $\Delta_v$ is therefore related to the variance of $\eta$ through
\begin{equation}
  \left( \frac {2 \omega_{\rm D}} {\Delta t} - \omega_D^2 \right)  \langle \Delta_v^2 \rangle 
  			= \langle \eta^2 \rangle
			\; \; ,
\end{equation}
since $\langle \Delta_v \eta \rangle = 0$ for Markovian impulses. In the continuous limit ($\Delta t \rightarrow 0$), this result implies that driving with $\langle \eta(t) \eta(t - \tau) \rangle = 2 \omega_{\rm D} (\sigma_{\rm t} / \sigma_{\rm v})^2 \delta(\tau)$ produces an rms velocity dispersion $\langle \Delta_v^2 \rangle^{1/2} = \sigma_{\rm t} / \sigma_{\rm v}$, and the numerical models of \S 3 converge to this result in the limit $\omega_{\rm D} \Delta t \ll 1$ when gravity is negligible.  However, the relationship between $\langle \eta^2 \rangle$ and $\langle \Delta_v^2 \rangle$ is more complex when random driving couples with thermally unstable gravity waves, as shown in \S 3.

\section{Comparing Cooling, Mixing, and Buoyancy}
\label{sec-AppendixMixing}

Numerical simulations show that damping of buoyant oscillations in an otherwise quiescent medium with $t_{\rm cool} \gg t_{\rm ff}$ causes thermal instability to saturate before it can produce condensation \citep[e.g.,][]{McCourt+2012MNRAS.419.3319M,Meece_2015ApJ...808...43M,ChoudhurySharma_2016MNRAS.457.2554C}.  Simulations also show that condensation can proceed in a medium with $t_{\rm cool} /t_{\rm ff} \approx 10$ in which driving of turbulence produces a velocity dispersion $\sigma_{\rm t} \sim 0.5 \sigma_v$ \citep[e.g.,][]{Gaspari+2013MNRAS.432.3401G}.  However, too much turbulence causes mixing that inhibits condensation if the timescale for mixing is much shorter than the cooling time \citep[e.g.,][]{BanerjeeSharma_2014MNRAS.443..687B,Gaspari_2017MNRAS.466..677G}.  Turbulence can also raise $t_{\rm cool}$ if the turbulent energy dissipation rate exceeds the cooling rate \citep[e.g.,][]{Gaspari+2013MNRAS.432.3401G}.

A recent analysis of cluster cores by \citet{Gaspari_2018ApJ...854..167G} posits that the condition for condensation in a turbulent medium should be $t_{\rm cool} / t_{\rm eddy} \approx 1$, with 
\begin{equation}
  t_{\rm eddy} = \frac {2 \pi r} {\sigma_{\rm t}} \left( \frac {L} {r} \right)^{1/3}
  \; \; .
\end{equation} 
The quantity $L$ specifies the length scale on which turbulence with a velocity dispersion $\sigma_{\rm t}$ is injected.  A turbulent cascade then produces a velocity dispersion $(r/L)^{1/3} \sigma_{\rm t}$ at radii $r < L$, and the corresponding eddy turnover time is assumed to be the time required to travel a distance $2 \pi r$ at that speed.

The \citet{Gaspari_2018ApJ...854..167G} condensation criterion can be restated in terms of $t_{\rm cool} / t_{\rm ff}$ as
\begin{equation}
  \frac {t_{\rm cool}} {t_{\rm ff}} \approx  2 \pi \frac {\sigma_v} {\sigma_{\rm t}} 
  							\left( \frac {r} {L} \right)^{1/3}
			\; \; .
\end{equation} 
\citet{Gaspari_2018ApJ...854..167G} show that this condition is consistent with observations of the turbulent velocity dispersions and radii over which multiphase gas extends in galaxy cluster cores.  However, it is indistinguishable from the criterion $t_{\rm cool} / t_{\rm ff} \approx 10$--15 for $\sigma_{\rm t} \approx 0.5 \sigma_v$, which \S 3 of this paper derives from buoyancy considerations.  

Clearly, $t_{\rm cool} / t_{\rm eddy} \approx 1$ cannot be the only criterion that matters, because reducing $\sigma_{\rm t}$ would then allow condensation to happen at arbitrarily large values of $t_{\rm cool} / t_{\rm ff}$.  The numerical experiments in this paper show that $\sigma_{\rm t}$ also needs to be large enough to suppress damping of entropy perturbations through dissipation of buoyant oscillations.  Those experiments account for mixing through the parameter $f_{\rm mix} = \omega_{\rm mix} / \omega_{\rm D}$.  Setting $\omega_{\rm mix} = t_{\rm eddy}^{-1}$ gives
\begin{equation}
  f_{\rm mix} \approx \frac {1} {2 \pi} \left( \frac {r} {L} \right)^{1/3}
\end{equation}
for $\omega_{\rm D} \approx (\sigma_{\rm t} / \sigma_{\rm v}) t_{\rm ff}^{-1}$, which is the appropriate damping rate for eddies of size $\approx r$.  This result implies that assuming $0.1 \lesssim f_{\rm mix} \lesssim 0.2$, as in Figure 7, implicitly approximates the damping effects of mixing that motivate the \citet{Gaspari_2018ApJ...854..167G} criterion.  According to the numerical trials represented in the top panel of Figure 7, buoyancy and mixing together limit condensation to a region in the $\sigma_{\rm t} / \sigma_v$--$t_{\rm cool}/t_{\rm ff}$ plane below the locus with $t_{\rm cool}/t_{\rm ff} \approx 25 (\sigma_{\rm t} / \sigma_v)$.

\section{Global vs.~Local Condensation Criteria}

Equation (20) is a criterion for condensation that implicitly specifies the critical Eulerian entropy contrast factor $\delta_c$ below which a perturbation inevitably condenses in terms of the ambient (global) value of $t_{\rm cool}/t_{\rm ff}$.  The local cooling time of gas with that Eulerian entropy contrast is $e^{\alpha_{\rm ti} \delta_K}$ times the ambient value of $t_{\rm cool}$, according to Equation (15).  Therefore, Equation (20) implies that condensation of a perturbation becomes inevitable if its {\em local} ratio of $t_{\rm cool} / t_{\rm ff}$ is
\begin{equation}
  \left( \frac {t_{\rm cool}} {t_{\rm ff}} \right)_{\rm local} \lesssim
     \frac {\alpha_{\rm ti}} {\alpha_K} 
     \left[ \frac {kr} {2} \left| \delta_c \right| \max \left( \frac {5} {3} , | \delta_c | \right) \right]^{1/2}
     \; \; .
\end{equation}
The local criterion depends on $kr$ because perturbations with large $kr$ are more susceptible to drag and have smaller terminal velocities that reduce the damping effects of buoyancy.  If both $kr$ and $| \delta_c |$ are of order unity, then the local value of $t_{\rm cool} / t_{\rm ff}$ at which condensation becomes inevitable is $(t_{\rm cool}/t_{\rm ff})_{\rm local} \approx \alpha_{\rm ti} / \alpha_K = 1.8$ for $\alpha_{\rm ti} = 6/5$ and $\alpha_K = 2/3$.

\section{Feedback-Induced Coupling between Turbulence and Condensation}

Equations (21) and (22) attempt to represent the qualitative features of coupling between precipitation, feedback, and turbulence at the center of a massive galaxy by reducing all the attendant complexity to just two degrees of freedom, $\sigma_{\rm t}$ and $t_{\rm cool}$, which respectively represent the 1D turbulent velocity dispersion and ambient cooling time of a single-zone system.  The specific condensation rate of ambient gas is expressed as $f_{\rm p} t_{\rm cool}^{-1}$, where $f_{\rm p}$ is a given function of $\sigma_{\rm t}$ and $t_{\rm cool}$.  Multiplying $f_{\rm p}t_{\rm cool}^{-1}$ by $\epsilon_H c^2$ gives the specific energy input rate coming from precipitation-fueled feedback.  

The system's specific energy in the form of turbulence is $3 \sigma_{\rm t}^2 / 2$, and it decays into heat at a rate $\Gamma \sigma_{\rm t} / \sigma_v t_{\rm ff}$, where $\Gamma$ is a tunable free parameter.  Typically, one expects $\Gamma \sim r / L$, where $L$ is the injection scale for turbulence with a velocity dispersion $\sigma_v$.  Given those assumptions, Equation (21) governs how $\sigma_v$ evolves in a system where a fraction $f_{\rm t}$ of feedback energy injection goes into turbulence before it dissipates into heat.

Equation (22) follows from the specific heating rate
\begin{equation}
  (1 - f_{\rm t}) \frac {f_{\rm p} \epsilon_H c^2}  {t_{\rm cool}}
  +  \left( \frac {3} {2} \sigma_{\rm t}^2 \right) \frac {\Gamma \sigma_{\rm t}} {t_{\rm ff} \sigma_v}
  \; \; ,
\end{equation}
in which the first term represents direct heating by feedback and the second represents heating via turbulent dissipation.  By design, the system's specific thermal energy remains constant at $3 \sigma_v^2$ because of gravitational work in the isothermal potential as its gas gains and loses entropy.  Consequently, the entropy equation for the gas is
\begin{equation}
  \frac {d \ln K} {dt} =   \frac {1} {t_{\rm heat}} - \frac {1} {t_{\rm cool}} = 
    (1 - f_{\rm t}) \frac {f_{\rm p} \epsilon_H c^2}  {3 \sigma_v^2 t_{\rm cool}}
    +  \frac {\Gamma} {2 t_{\rm ff}} \left( \frac {\sigma_{\rm t}} {\sigma_v} \right)^3 
    - \frac {1} {t_{\rm cool}}
  \; \; ,
\end{equation}
where $t_{\rm heat}$ is the specific thermal energy divided by the specific heating rate.   Application of the relation $d \ln t_{\rm cool} / d \ln K = 3/2$ for optically thin radiative losses from gas kept at constant temperature by gravitational work then gives 
\begin{equation}
  \frac {d \ln t_{\rm cool}} {dt} =    
    (1 - f_{\rm t}) \frac {f_{\rm p} \epsilon_H c^2}  {2 \sigma_v^2 t_{\rm cool}}
    +  \frac {3\Gamma} {4 t_{\rm ff}} \left( \frac {\sigma_{\rm t}} {\sigma_v} \right)^3 
    - \frac {3} {2 t_{\rm cool}}
  \; \; ,
\end{equation}
which multiplication by $t_{\rm cool}$ simplifies to Equation (22).

\bibliographystyle{apj}

\bibliographystyle{apj}

\end{document}